\documentstyle[12pt]{article}
\newcommand{\be}{\begin{equation}}
\newcommand{\ee}{\end{equation}}
\newcommand{\bea}{\begin{eqnarray}}
\newcommand{\eea}{\end{eqnarray}}
\newcommand{\beaa}{\begin{eqnarray*}}
\newcommand{\eeaa}{\end{eqnarray*}}
\newcommand{\ol}{\overline}

\newcommand{\al}{\alpha}
\newcommand{\bt}{\beta}
\newcommand{\th}{\theta}

\newcommand{\la}{\lambda}

\newcommand{\Ga}{\Gamma}
\newcommand{\si}{\sigma}
\newcommand{\om}{\omega}
\newcommand{\fo}{\frac{\om}2}
\newcommand{\rb}{\right]}
\newcommand{\lb}{\left[}
\newcommand{\lag}{\langle}
\newcommand{\rag}{\rangle}
\newcommand{\bl}{\biggl(}
\newcommand{\br}{\biggr)}
\newcommand{\cbl}{\biggl\{ }
\newcommand{\cbr}{\biggr\} }
\newcommand{\cR}{\check{R}}
\newcommand{\cP}{\check{P}}
\newcommand{\tm}{\tilde{\mu}}
\newcommand{\nn}{\nonumber}
\newcommand{\fns}{\footnotesize}
\newcommand{\scs}{\scriptsize}
\newcommand{\shs}{\shortstack}
\newcommand{\monoid}{\raisebox{-10pt}{\usebox{\monoidbox}}}
\newcommand{\identity}{\raisebox{-10pt}{\usebox{\identitybox}}}
\newcommand{\braid}{\raisebox{-10pt}{\usebox{\braidbox}}}
\newcommand{\antibraid}{\raisebox{-10pt}{\usebox{\antibraidbox}}}
\thicklines
\newsavebox{\identitybox}
\savebox{\identitybox}{
\begin{picture}(30,26)
\put(0,0){\line(1,0){26}}
\put(0,26){\line(1,0){26}}
\end{picture}}
\newsavebox{\monoidbox}
\savebox{\monoidbox}{
\begin{picture}(30,26)
\put(0,13){\oval(20,26)[r]}
\put(26,13){\oval(20,26)[l]}
\end{picture}}
\newsavebox{\braidbox}
\savebox{\braidbox}{
\begin{picture}(30,26)
\put(0,26){\line(1,-1){26}}
\put(0,0){\line(1,1){10}}
\put(26,26){\line(-1,-1){10}}
\end{picture}}
\newsavebox{\antibraidbox}
\savebox{\antibraidbox}{
\begin{picture}(30,26)
\put(0,0){\line(1,1){26}}
\put(0,26){\line(1,-1){10}}
\put(26,0){\line(-1,1){10}}
\end{picture}}
\advance\textheight by \topskip
\begin{document}
\def\ss{\scriptstyle}
\baselineskip 17pt
\parindent 10pt
\parskip 15pt

\newpage
\begin{titlepage}
\begin{flushright}
DAMTP-95-33\\
hep-th/9506169\\
revised September 95
\end{flushright}
\vspace{1cm}
\begin{center}
{\Large {\bf Exact S-matrices for $d_{n+1}^{(2)}$ affine Toda
solitons \vspace{.2cm}\\
 and their bound states}}\\
\vspace{1.5cm}
{\large G.M. Gandenberger}\footnote{e-mail:
G.M.Gandenberger@damtp.cam.ac.uk}
and {\large N.J. MacKay}\footnote{e-mail: N.J.MacKay@damtp.cam.ac.uk}\\
\vspace{3mm}
{\em Department of Applied Mathematics and Theoretical Physics,}\\
{\em Cambridge University}\\
{\em Silver Street, Cambridge, CB3 9EW, U.K.}\\
\vspace{2cm} {\bf{ABSTRACT}}
\end{center}
\begin{quote}
We conjecture an exact S-matrix for the
scattering of solitons in $d_{n+1}^{(2)}$ affine Toda field
theory in terms of the R-matrix of the quantum group
$U_q(c_n^{(1)})$. From this we construct the scattering amplitudes for
all scalar bound states (breathers) of the theory.
This S-matrix conjecture is justified by detailed examination of its
pole structure. We show that a breather-particle identification holds
by comparing the S-matrix elements for the lowest breathers with the
S-matrix for the quantum particles in real affine Toda field theory,
and discuss the implications for various forms of duality.
\end{quote}

\vspace{1cm}
\begin{center}
To appear in NUCLEAR PHYSICS B
\end{center}

\vfill

\end{titlepage}

\section{Introduction}

The Lagrangian density of affine Toda field theory (ATFT)
with imaginary coupling-constant (`imaginary ATFT')
can be written in the form
\be\label{L}
{\cal L}={1\over2}(\partial_\mu \phi)(\partial^\mu\phi)+{m^2\over\bt
^2}\sum_{j=0}^n n_j(e^{i\bt\al_j\cdot\phi}-1)\;.
\ee
The field $\phi(x,t)$ is an $n$-dimensional vector, $n$ being the rank
of the finite Lie algebra $g$. The $\alpha_j$ ($j=1,...,n$) are
the simple roots of $g$; $\alpha_0$ is chosen such that the inner
products among the elements of the set
\{$\al_0,\al_j$\} are described by one of the extended Dynkin
diagrams of an affine algebra $\hat{g}$. It is expressible in terms
of the other roots by
$$
\al_0=-\sum_{j=1}^n n_j\alpha_j
$$
where the $n_j$ are positive integers, and $n_0=1$.
$\bt$ is a coupling constant (which in the notation here is a positive
real number) and $m$ a mass scale. We take the roots
to be normalised so that $|$longest root$|^2=2k$, where $k$ is
the twist of the affine algebra.

Four years ago Hollowood discovered classical soliton solutions
\cite{hollo92} to $a_n^{(1)}$ imaginary ATFTs. Solitons were
subsequently found for ATFTs based
on other algebras \cite{macka93}, and a general construction based on
vertex operators followed \cite{olive93}.
For algebras other than $a_1^{(1)}$
(when the ATFT is the sine-Gordon model), the Hamiltonian is complex;
yet the solitons have real energy and momenta, and real higher
conserved charges \cite{free94}. The solitons fall
into species labelled (for untwisted affine algebras) by spots
on the Dynkin diagram of $g$, and have
topological charges which lie in, but by no means fill,
the fundamental representations of $g$ \cite{mcg94}. (It has
to be hypothesised that in the quantum theory states exist
which fill the quantized affine algebra representations.) Further,
like the sine-Gordon model, these theories have continuous spectra
of bound states with zero topological charge (`breathers') of each
species; but, unlike the sine-Gordon model, there are also similar
states with non-zero charge (`excited solitons' or `breathing
solitons') \cite{harde94}.

Using the semiclassical methods of \cite{dashen77} the order $\bt^2$
quantum mass corrections have been calculated
\cite{deliu94b,hollo93b,macka94} and
the semiclassical limit of the $S$-matrix computed \cite{spence95}.
However, the $\hat{g}$ ATFT has as a dynamical symmetry
an underlying $U_q(\hat{g}^\vee)$ charge algebra \cite{berna90,feld91}
(where the superscript denotes the dual of $\hat{g}$,
{\em i.e.\ }the algebra obtained by replacing all the roots
with co-roots, $\al\mapsto {2\over\al^2}\al$), and
conservation of these charges allows us to solve for the exact
soliton $S$-matrix, which satisfies the Yang-Baxter equation (YBE).
This was done for the $a_n$ case in \cite{hollo93}.

In the quantum theory the mass spectrum of the excitations
becomes discrete, and in the sine-Gordon theory it is
natural to identify the lowest breather state with the
quantum particle (the quantized vacuum excitation). This seems to
apply more generally: comparing with the mass and S-matrix
calculations for the particles carried out in \cite{brade90,chris90}
for the real coupling case ($\bt$ purely imaginary in (\ref{L}), which
we shall call `real ATFT'\footnote{For a recent review of real ATFT
see \cite{corri94}.}) we find that, when $\hat{g}$ is self-dual, the
lowest breather masses implied by the semiclassical soliton mass
corrections and the exact S-matrix poles are precisely those of the
particles \cite{hollo93b,macka94}. Further, in the only case
investigated in detail ($a_2^{(1)}$ in \cite{gande95}) the $S$-matrix
also matches.

The theories based on self-dual affine algebras (the simply-laced
algebras plus $a_{2n}^{(2)}$) are the least interesting, in that
the classical mass ratios for solitons, breathers and particles
are unaltered by quantum mass corrections. In contrast, for
nonsimply-laced $\hat{g}$ the relations among the masses and
$S$-matrices of these objects are little understood.
The nonsimply-laced $\hat{g}$ fall into two categories, both of which
may be obtained from simply-laced theories as subspaces
invariant under automorphisms (`foldings') of the extended Dynkin
diagram. For the twisted algebras ({\em i.e.\ }where the automorphism
involves the extended root) both particles and solitons
are a subset of those of the parent theory. Their
masses, however, now renormalise differently, although the
corrections are in the same ratio for solitons and particles
of the same species. Any attempt to construct exact $S$-matrices
must take account of this fact via the introduction of flexible
pole structure, the possibility of which has only recently
been recognised \cite{babic94}. The least tractable case is
that of the untwisted nonsimply-laced algebras ({\em i.e. }where
the automorphism does not involve the extended root), where
even the classical masses of the solitons and particles
are not proportional: in fact, the masses of the solitons
of the ATFT based on (the affine extension of) $g$ are proportional
to those of the particles of the $g^\vee$ ATFT \cite{olive93}, a fact
suggestive of some kind of `Lie duality' (in contrast to the
`affine duality' found \cite{brade90,kausch90} between the particles
of the $\hat{g}$ real ATFT in the weak-coupling regime and
those of the $\hat{g}^\vee$ theory in the strong regime).
However, this can only be made to work in a rather
subtle way in the quantum theory: if we assume
that the ratios of the quantum masses to the classical masses of solitons
and particles of the same species are independent of the species
(even for untwisted nonsimply-laced $\hat{g}$, where the
ratios of the classical masses depend on the species)\footnote{
See \cite{deliu94b,macka94}
and the discussions therein: this assumption will hold
only if the na\"{\i}ve semiclassical approach fails.}, then
this duality persists only between imaginary-coupled solitons
and real-coupled particles.

It is only feasible to construct exact $S$-matrices where
the spectral decomposition of the $U_q(\hat{g}^\vee)$-invariant
solutions of the YBE (`$R$-matrices') is known for all species.
This is only
the case for those algebras for which the fundamental
representations of $U_q(\hat{g}^\vee)$ are irreducible
as representations of the Lie subalgebra; and this is only true
where the Lie algebra is $a_n$ or $c_n$. The $a_n$ case
has been investigated by Hollowood \cite{hollo93}, who examined
the soliton $S$-matrices but did not fuse them to
obtain breather or excited soliton \mbox{$S$-matrices}, which
are in some ways rather subtle because of the non-self-conjugacy
of the particles and solitons. The $a_2^{(1)}$ case has been
investigated in detail \cite{gande95}. In this paper we
investigate the $d_{n+1}^{(2)}$ ATFTs, which therefore have
$U_q(c_n^{(1)})$ symmetry and $U_q(c_n)$-invariant $S$-matrices.
We construct soliton-soliton $S$-matrices, and from them
construct the breather-soliton, breather-breather and
breather-excited soliton $S$-matrices, although the excited
soliton-excited soliton $S$-matrices remain beyond our scope.

In section two we gather together some necessary facts about
$U_q(c_n)$-invariant $R$-matrices, and in section three
we fuse the soliton $S$-matrices to obtain $S$-matrices
for the breather bound states, finding that those for
the lowest breather states are precisely those for the particles
\cite{deliu92}, supporting the identification of these objects.
In section four we investigate the soliton $S$-matrices' pole
structure and present what we believe to be the minimal set of
three-point couplings necessary for the bootstrap to close.

This paper, dealing with a twisted theory, should be seen as an
intermediate step between the simply-laced theories and the untwisted
nonsimply-laced theories, which will be the subject of future work.
In section five we expand on our discussion above and present
a general scheme for the investigation of these theories
and of affine and Lie duality. Four appendices deal, respectively,
with generalisation to other algebras, details of crossing symmetry
for the basic $R$-matrix,
details necessary for the calculation of the scalar factor in
the $S$-matrix, and the pole structure of the (rational)
$c_n$-invariant $S$-matrices.

\newpage
\section{$U_q(c_n)$-invariant $R$-matrices}

The $U_q(c_n)$-invariant $R$-matrices can be described
using the tensor product graph \cite{macka91,zhang91} and are given by
\cite{deliu94,hollo94,macka92}
\be
\cR_{a,b}^{(TPG)}(x) = \sum_{p=0}^{min(b,n-a)} \sum_{r=0}^{b-p}
\prod_{i=1}^p \lag a-b+2i \rag \prod_{j=1}^r \lag 2n+2-a-b+2j \rag
\cP_{\la_{a+p-r}+\la_{b-p-r}} \label{specdecom}
\ee
in which $a,b=1,..,n$; $a\geq b$ and $\lag k \rag =
\frac{1-xq^k}{x-q^k}$. $\cP_{\la}$ denotes the projector onto the
($q$-deformation of the) module of the irreducible $c_n$-representation
with highest weight $\la$ (which we shall denote $V_i$ when
$\la=\la_i$,
the $i$th fundamental weight; $\la_0\equiv 0$). $\cR_{a,b}^{TPG}(x)$
acts as an intertwiner on these modules
\[
\cR_{a,b}^{(TPG)}(x): V_a\otimes V_b \to V_b\otimes V_a \;.
\]
The tensor product graph (TPG) itself,
in which the coefficients of two linked representations in the graph
are in the ratio $\lag \Delta \rag$ where $\Delta$ is the difference
in the two values of the quadratic Casimir operator, is
$$
\begin{array}{ccccccccccc}
\lambda_a+\lambda_b & \rightarrow & \lambda_{a+1}+\lambda_{b-1}&
\cdots & \rightarrow & \lambda_n+\lambda_{a+b-n} &  \cdots &
\rightarrow & \lambda_{a+b-1}+\lambda_1 & \rightarrow &
\lambda_{a+b}\\[0.1in]
\downarrow & & \downarrow & & & \downarrow & & & \downarrow \\[0.1in]
\lambda_{a-1}+\lambda_{b-1} & \rightarrow & \lambda_a+\lambda_{b-2} &
\cdots & \rightarrow & \lambda_{n-1}+\lambda_{a+b-n-1} &  \cdots &
\rightarrow & \lambda_{a+b-2}\\[0.1in]
\vdots & & \vdots & & & \vdots \\[0.05in]
\downarrow & & \downarrow & & & \downarrow \\[0.1in]
\lambda_{n-b} + \lambda_{n-a}& \rightarrow &
\lambda_{n-b+1}+\lambda_{n-a-1}&  \cdots & \rightarrow
& \lambda_{2n-a-b} \\[0.1in]
\vdots & & \vdots \\[0.05in]
\downarrow & & \downarrow \\[0.1in]
\lambda_{a-b+1}+\lambda_1 & \rightarrow & \lambda_{a-b+2} \\[0.1in]
\downarrow \\[0.1in]
\lambda_{a-b} \\[0.1in]
\end{array}
$$
For $a+b>n$ the graph truncates at the $(n-a+1)$th column, since the
representations to the right of this column in the graph are then
no longer present in the decomposition of $V_a\otimes V_b$. This fact
will be crucial in examining the orders of the poles.

We must now decide the dependence of $x(\th,\bt)$ and $q(\bt)$
on the rapidity difference \mbox{$\th=\th_1-\th_2$} (where the
incident particles' rapidities are defined by $p_i=(m_i\cosh\th_i,
m_i\sinh\th_i)\;$) and on $\bt$. For the $d_{n+1}^{(2)}$ case, and
in order to give the correct soliton mass ratios, we take
$$
x= e^{(n+1)\lambda\th}\;, \hspace{0.3in}{\rm where}\;\;\;
\lambda = {4\pi\over \bt^2} - {2n\over n+1}\;,
$$
and
\be\label{xqdef}
q=e^{\om i\pi}\;, \hspace{0.3in}{\rm where}\;\;\;
\om={2\pi\over \beta^2}-1 \;.
\ee
In this paper we only discuss the case of generic $q$ ({\em i.e.\ }$q$
not a root of unity, equivalent to requiring that $\om$ not be
rational). For other algebras, as discussed in \cite{deliu95}, we
need the forms given in appendix \ref{data}, which are not yet
properly understood. It is for this reason that we cannot
yet give a derivation of the general $S$-matrix direct from
the charge algebra, which we leave for future investigation.

\subsection{Fusion and crossing properties}

We now proceed with some results which we shall need later
in fusing the $S$-matrices. First, we examine the fusion
properties of the $\cR^{(TPG)}$. We are always free to
rescale $R$-matrices by a scalar factor, and it turns out
that the $R$-matrices preserved by fusion are not those
given above, in which $\cP_{\la_a+\la_b}$ has coefficient $1$,
but those for which $\cP_{\la_{a+b}}$ has coefficient $1$.
Denoting these $R$-matrices by $\cR'$, we find
$$
\cR'_{a,b}(x) \equiv \prod_{j=1}^a \prod_{k=1}^b
\lb \cR'_{1,1}(xq^{-2-a-b+2j+2k}) \rb_{a+1-j,a+k}
$$
where the equation acts on $V_a\otimes V_b \subset V_1^{\otimes (a+b)}$
and $[\;]_{i,j}$ indicates that the $R$-matrix is taken to act
on the $i$th and $j$th $V_1$s. The product is taken in order of
increasing $j$ and $k$. The result holds because when $x=q^{a+b}$
anti-symmetrization now takes place on {\em all} spaces, and
$\cR_{a,b}$ projects onto $V_{a+b}$ with a coefficient which
must, by unitarity, equal $1$. This result still applies
when the TPG is truncated.

Using
$$
\cR^{(TPG)}_{a,b}(x) = \prod_{k=1}^b \,- \lag a-b+2k\rag \;
\cR'_{a,b}(x)
$$
we now obtain
\be\label{fusion}
k_{a,b}(x) \cR^{(TPG)}_{a,b}(x) \equiv \prod_{j=1}^a \prod_{k=1}^b
\lb \cR^{(TPG)}_{1,1}(xq^{-2-a-b+2j+2k}) \rb_{a+1-j,a+k}
\ee
where
\be
k_{a,b}(x) = \prod_{l=1}^b \,\prod_{m=1}^{a-1} \,-\lag
2l-2m+a-b\rag \;. \label{kab}
\ee

The second result we need is an explicit formula for $R$-matrix
crossing symmetry analogous to that conjectured to hold
for the $a_n$ case in (3.16) of \cite{hollo90}.
In appendix \ref{cross} we prove the result (\ref{cr11},\ref{c11}),
which we now generalise to
\be\label{crab}
c_{a,b}(i\pi-\th)\cR_{a,b}^{(TPG)cross}(x(i\pi-\th)) =
c_{a,b}(\th)\cR_{a,b}^{(TPG)}(x(\th))
\ee
in which
\be\label{cab}
c_{a,b}(\th) =  \prod_{k=1}^b
\sin(\pi(\mu-\fo(a-b+2k)))
\sin(\pi(\mu-\fo(2n+2-a-b+2k)))\;,
\ee
where
\be
\mu = -i\frac{(n+1)\la}{2\pi}\th \;. \label{mu}
\ee
The proof uses (\ref{fusion}) and is equivalent to showing
that the set of poles in the fused $\cR_{1,1}^{(TPG)}$,
minus the set of poles in $\cR_{a,b}^{(TPG)}$ (including those
which would appear in the truncated part of the TPG), is invariant
under $\th\mapsto i\pi-\th$. (It should be noted that $c_{a,b}$
has zeros exactly where $\cR^{(TPG)}_{a,b}$ has poles.)

The last result is one we shall need in order to calculate
the breather-soliton $S$-matrices in the next section,
\bea\nn
\cR^{(TPG)}_{a,0_b}(x(\th)) & \equiv &  P_0\otimes I_a\,.\,
I_b\otimes \cR^{(TPG)}_{a,b}(xq^{n+1}) \,.\,
\cR^{(TPG)}_{a,b}(xq^{-(n+1)})\otimes I_b \\ \label{sol-br}
& = & \frac{c_{a,b}(\th+\frac{i\pi}2-\frac{i\pi}{(n+1)\la})}
{c_{a,b}(-\th+\frac{i\pi}2-\frac{i\pi}{(n+1)\la})}
\; \cR^{(TPG)}_{b,a}(x^{-1}q^{n+1})\,.\,
\cR^{(TPG)}_{a,b}(xq^{-(n+1)}) \\\nn
& = & \prod_{k=1}^b
{\sin(\pi(\mu-\fo(a-b+2k-n-1)))
\sin(\pi(\mu-\fo(n+1-a-b+2k))) \over
\sin(\pi(\mu+\fo(a-b+2k-n-1)))
\sin(\pi(\mu+\fo(n+1-a-b+2k)))}\times I_a \;.
\eea
in which $I_a$ denotes the identity on $V_a$.
This result is made most evident in the following diagram:

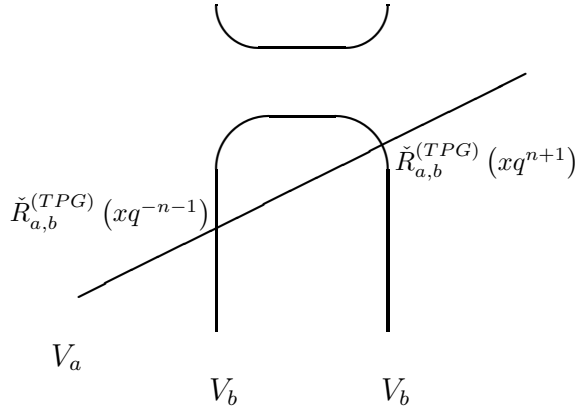
\begin{picture}(150,150)(-140,0)
\setlength{\unitlength}{1.3pt}
\put(70,25){\oval(50,125)[t]}
\put(70,120){\oval(50,25)[b]}
\put(5,35){\line(2,1){130}}
\put(-3,15){$V_a$}
\put(43,5){$V_b$}
\put(93,5){$V_b$}
\put(-15,57){\shs{\fns{$\cR_{a,b}^{(TPG)}
\left(xq^{-n-1}\right)$}}}
\put(97,72){\shs{\fns{$\cR_{a,b}^{(TPG)}
\left(xq^{n+1}\right)$}}}
\end{picture}

The first line acts on $V_a\otimes V_b \otimes V_b$, while the
second acts on $V_a\otimes V_b$, with  $\cR_{b,a}^{(TPG)}$
now acting in the crossed channel. Unitarity of the $R$-matrix
then gives the third line, a simple scalar factor acting on $V_a$
in the direct channel.

\section{Exact S-matrices for $d_{n+1}^{(2)}$ quantum affine Toda
field theory}
We now define the S-matrix for the scattering of elementary solitons
to be
\be
S_{a,b}(\th) = F_{a,b}(\mu(\th)) k_{a,b}(\th)\:
\tau_{21} \cR_{a,b}^{(TPG)}(\th) \tau_{12}^{-1}
\label{smatrix}
\ee
in which $\tau$ is the transformation from the homogeneous to
the spin gradation described in appendix \ref{data}, $k_{a,b}(\th)$ is
given by (\ref{kab}) and the overall scalar factor $F_{a,b}(\mu(\th))$
will be constructed below.
\subsection{Scalar factors}

The scalar factors $F_{a,b}(\mu)$ have to be chosen such that the
S-matrix $S_{a,b}(\th)$ satisfies the axioms of exact S-matrix theory
(for a more detailed account of the exact S-matrix axioms see, for
instance, \cite{hollo93,eden66,zamol79}).
The two axioms which determine the scalar factor up to a so-called
CDD-factor are the requirements of unitarity:
\be
S_{a,b}(\th)S_{b,a}(-\th) = I_b \otimes I_a \label{unit}
\ee
and crossing symmetry:
\be
S_{\ol a,b}(\th) = (I_b \otimes C_a)[\si
S_{b,a}(i\pi-\th)]^{t_2}\si(C_{\ol a}\otimes I_b) \;, \label{crossing}
\ee
in which $\ol a$ represents the conjugate of $a$, $C_a$ is the
charge conjugation operator, $t_2$ indicates transposition
in the second space and $\si$ is the permutation operator. In
the case of $d_{n+1}^{(2)}$ ATFT, in which all particles
are self-conjugate, (\ref{crossing}) reduces to
\be
S_{a,b}(\th) = [\si S_{b,a}(i\pi-\th)]^{t_2}\si \;.
\ee
In the following we will use these equations to obtain the
scalar factor $F_{1,1}$ in terms of products of gamma functions.
In order to exploit the crossing symmetry requirement we need to
examine the crossing symmetry properties of the R-matrix itself. As
mentioned in section 2.1 we
seek to find a scalar function $c_{1,1}(\th)$ such that
$c_{1,1}(\th)\cR_{1,1}^{(TPG)}(\th)$ is crossing symmetric,
i.e. satisfies equation (\ref{crab}). By expressing the R-matrix
in terms of the Birman-Wenzl-Murakami
algebra \cite{jimbo89,macka92b} we show in Appendix \ref{cross}
that this function is given by
\[
c_{1,1}(\th) = \sin(\pi(\mu - \om)) \sin(\pi(\mu -
(n+1)\om)) \;.
\]
Writing $F_{1,1}(\mu(\th)) = c_{1,1}(\th)f_{1,1}(\mu(\th))$ and using
equation (\ref{crossing}) for $a=b=1$ (noting that $\mu(i\pi-\th) =
-\mu+\frac{n+1}2\la$) we obtain the first condition\footnote{The
disagreement of this condition with equation (2.45) in
\cite{deliu95} is due to the fact that Delius' scalar factor is
different from the one we use, since his factor $c_{1,1}$ is
determined by the definition \mbox{$\cR_{1,1}(\th) =
(\pi_1^{(\th)}\otimes\pi_1^{(0)}) {\cal R}$} and is therefore
not equal to our $c_{1,1}(\th)$.}
\be
f_{1,1}(-\mu+\frac{n+1}2\la) = f_{1,1}(\mu)\;. \label{iter1}
\ee
Using $\lag l \rag_{(-\th)} = 1/\lag l \rag_{(\th)}$ and the
spectral decomposition of the R-matrix (\ref{specdecom})
\[
\cR^{(TPG)}_{1,1}(\th) = \cP_{2\la_1} + \lag 2\rag \cP_{\la_2} + \lag
2n+2\rag \cP_0  \label{specdecom1}
\]
we can see immediately that
\[
\cR_{1,1}^{(TPG)}(\th) \cR_{1,1}^{(TPG)}(-\th) = I_1 \otimes I_1 \;.
\]
{}From this and equation (\ref{unit}) we obtain the second condition
\be
f_{1,1}(\mu)f_{1,1}(-\mu) =
c_{1,1}^{-1}(\th)c_{1,1}^{-1}(-\th) \;. \label{iter2}
\ee
The standard method (see for instance \cite{deliu91,shank78})
for solving (\ref{iter1}) and (\ref{iter2})
consists of choosing a suitable starting function $f^{(1)}(\mu)$,
which satisfies
equation (\ref{iter2}), and multiplying it by a factor such that equation
(\ref{iter1}) is satisfied. Then equation (\ref{iter2}) is violated
again and one has to multiply this by another factor. Continuing this
iteration process one ends up with an infinite product
that satisfies both equations.
The general solution to equations (\ref{iter1}) and (\ref{iter2}) is
given by
\be
f_{1,1}(\mu) = \prod_{j=1}^{\infty} \frac{f^{(1)}[\mu+(n+1)\la(j-1)]
f^{(1)}[-\mu+(n+1)\la(j-\frac12)]}
{f^{(1)}[\mu+(n+1)\la(j-\frac12)]
f^{(1)}[-\mu+(n+1)\la j]} \label{itersol}
\ee
for any function $f^{(1)}(\mu)$ with
$f^{(1)}(\mu)f^{(1)}(-\mu) = c_{1,1}^{-1}(\th)c_{1,1}^{-1}(-\th)$.
In order to choose an appropriate starting function $f^{(1)}(\mu)$ we
rewrite
\bea
c_{1,1}^{-1}(\th)c_{1,1}^{-1}(-\th) &=& \frac1{\pi^4} \Ga(\mu-\om)
\Ga(1-\mu+\om) \Ga(\mu-(n+1)\om) \Ga(1-\mu+(n+1)\om) \nn \\
&&\times \Ga(-\mu-\om) \Ga(1+\mu+\om) \Ga(-\mu-(n+1)\om)
\Ga(1+\mu+(n+1)\om) \;. \nn
\eea
For our purpose the only appropriate combination of gamma functions as
a starting function is
\be
f^{(1)}(\mu) = \frac1{\pi^2} \Ga(\mu-\om) \Ga(\mu-(n+1)\om)
\Ga(1+\mu+\om) \Ga(1+\mu+(n+1)\om) \;.
\ee
(Using any other combination will lead to a scalar factor with an
infinite number of poles on the physical strip.)
Using the solution (\ref{itersol}) and $\la = 2\om +\frac2{n+1}$ we
finally obtain the following scalar factor:
\bea
F_{1,1}(\mu) &=& \prod_{j=1}^{\infty} \frac{\Ga(\mu+(n+1)\la j-\om)
\Ga(\mu+(n+1)\la j-(2n+1)\om-1)}{\Ga(-\mu+(n+1)\la j-\om)
\Ga(-\mu+(n+1)\la j-(2n+1)\om-1)} \nn \\
&&\times \frac{\Ga(\mu+(n+1)\la j-(n+1)\om) \Ga(\mu+(n+1)\la
j-(n+1)\om-1)} {\Ga(-\mu+(n+1)\la j-(n+1)\om) \Ga(-\mu+(n+1)\la
j-(n+1)\om-1)} \nn \\
&&\times
\frac{\Ga(-\mu+(n+1)\la j-(n+2)\om-1) \Ga(-\mu+(n+1)\la j-n\om)}
{\Ga(\mu+(n+1)\la j-(n+2)\om-1) \Ga(\mu+(n+1)\la j-n\om)} \nn \\
&&\times \frac{\Ga(-\mu+(n+1)\la j-2(n+1)\om-1) \Ga(-\mu+(n+1)\la j)}
{\Ga(\mu+(n+1)\la j-2(n+1)\om-1) \Ga(\mu+(n+1)\la j)} \;. \label{F11}
\eea

The bootstrap principle of exact S-matrix theory can be applied to
obtain the general scalar factor $F_{a,b}(\mu)$. We will show this
in detail only for $F_{2,1}$. After a careful study of the
pole structure of $S_{1,1}(\th)$ it emerges that $S_{1,1}(\th)$
projects onto the module $V_2$ at the following poles (written in
terms of $\mu$ which was defined in (\ref{mu})):
\[
\mu(\th_p^{(2)}) = \om-p
\hspace{1cm} (\mbox{for } p = 0,1,...\leq \om).
\]
The lowest of these poles $(p=0)$ corresponds to the fusion process
$A^{(1)}+A^{(1)} \to A^{(2)}$ and from the bootstrap principle of
analytic S-matrix theory we know
\be
S_{2,1}(\th) = \lb S_{1,1}\left(\th+\frac{\th_0^{(2)}}2\right)\rb_{1,3}
\lb S_{1,1}\left(\th-\frac{\th_0^{(2)}}2\right)\rb_{1,2}.
\ee
Since we also know from equation (\ref{fusion}) that
$\lb\cR^{(TPG)}_{1,1}(\th+\frac{\th_0^{(2)}}2)\rb_{1,2}
\lb\cR^{(TPG)}_{1,1}(\th-\frac{\th_0^{(2)}}2)\rb_{1,3} = k_{2,1}(\th)
\cR^{(TPG)}_{2,1}(\th)$
we obtain the result:
\[
F_{2,1}(\mu) = F_{1,1}(\mu+\fo)  F_{1,1}(\mu-\fo) \;.
\]
Continuing this fusion procedure we find in general that the pole at
\be
\mu(\th_0^{(a+b)}) = \frac{a+b}2 \om \label{lowpole}
\ee
corresponds to the fusion process $A^{(a)}+A^{(b)}=A^{(a+b)}$ of
fundamental solitons and therefore we obtain the following expression for
the general scalar factor:
\be
F_{a,b}(\mu) = \prod_{j=1}^a \prod_{k=1}^b F_{1,1}(\mu +
\fo(2j+2k-a-b-2)) \;. \label{Fab}
\ee

\subsection{Soliton masses}

We conjecture that the S-matrix defined in
(\ref{smatrix}) describes the scattering of solitons in
$d_{n+1}^{(2)}$ ATFT. We explain how we expect the direct channel
simple poles to match the soliton masses for general algebras
in appendix \ref{data}; here we point out
that the lowest poles (\ref{lowpole}) which are at $x(\th^{(a+b)}_0) =
q^{a+b}$ match precisely the soliton mass ratios calculated in
\cite{macka94}. We will demonstrate this briefly.
We know that the classical soliton masses for $d_{n+1}^{(2)}$ ATFT (in
which $h = n+1$) are given by
\be
M_a^{Cl} = 8\sqrt{2}\frac{hm}{\beta^2}\sin\left(\frac{a\pi}{2h}\right) \;.
\ee
Via the usual mass relation
\be
M^2_{a+b} = M^2_a + M^2_b + 2 M_a M_b \cosh(\th^{(a+b)}_0)  \label{massform}
\ee
the poles (\ref{lowpole}) determine the quantum soliton masses (up to
an overall scale factor $C$) to be
\be
M_a=C8\sqrt{2}\frac{hm}{\beta^2}\sin\left(
\frac{a\pi}{h}\left(\frac12-\frac1{h\la}\right)\right)\;.
\label{qumass}
\ee
Expanding this in terms of $\beta^2$ we obtain
\[
M_a =
C8\sqrt{2}\frac{hm}{\beta^2}\sin\left(\frac{a\pi}{2h}\right
)[1-\beta^2\frac{a}{4h^2}
\cot\left(\frac{a\pi}{2h}\right)] + O(\beta^2)
\]
which coincides with the result in \cite{macka94} (table 4) provided that
\bea
C &=& 1 + \frac{\beta^2}{8h}\cot\left(\frac{\pi}{2h}\right)
- \frac{\beta^2}{4\pi}
\frac{h^{\vee}}{h} + O(\beta^4) \nn \\
&=& \frac{\beta^2}{4\pi}\la + \frac{\beta^2}{8h}\cot\left(
\frac{\pi}{2h}\right) + O(\beta^4) \;.
\eea
(It is not yet clear to us why the scale factor $C$ should have
this form.)
We can also see that (up to an overall scale factor)
the change from the classical to the quantum masses corresponds to a
shift of the Coxeter number $h$ to a so-called quantum Coxeter number:
\[ h \to h+\frac1{\om}\;. \]
This is similar to the situation in real ATFT \cite{deliu92} and has
also been pointed out in \cite{deliu95}. For a general discussion
see appendix \ref{data}.

We also should remark on the fact that our conjecture of the
soliton S-matrix does not contain any additional CDD factors. In
\cite{hollo93} a minimal Toda factor was necessary for the
construction of a consistent S-matrix for $a_n^{(1)}$ affine Toda
solitons. In the case of non-simply laced ATFTs minimal
(coupling constant independent) S-matrices do not exist (see
\cite{deliu92}). However, we will see in the following sections that
the scalar factor
defined above already contains all the poles and zeros on the physical
strip (i.e. $0\leq \mbox{Im}(\th) \leq \pi$) necessary to satisfy the
bootstrap equations, and we do not need to include any additional CDD
factor.
In the following sections we will further justify this
conjecture by comparing the S-matrix elements of scalar bound states
with the results in real ATFT and describing the pole structure in more
detail.

\subsection{Breather S-matrices}

By applying the fusion procedure (or bootstrap method) in this
section we construct the \mbox{S-matrix} elements for the scattering
of bound states. Here this procedure is applied directly to the full
S-matrix $S_{a,b}(\th):V_a\otimes V_b \to V_b\otimes V_a$, whereas in
\cite{gande95} the same method  has been applied to the scattering
amplitudes (which are just scalar functions) of single solitons by
using the Zamolodchikov algebra formalism.

Projectors onto singlets, which correspond to scalar bound states of
elementary solitons, only appear in the elements
$\cR_{a,a}^{(TPG)}$ for $a=1,2,...,n$.
We will call these scalar bound states `breathers' and denote them by
$B^{(a)}_p$. The poles corresponding
to these breathers must be in the prefactor $\lag 2n+2 \rag$, since
this is the factor which in the spectral decomposition
(\ref{specdecom}) appears in front of $\cP_0$ exclusively.
$\lag 2n+2 \rag$ has the following poles on the physical strip:
\be
\mu = (n+1)\om +1-p \hspace{1cm} \mbox{(for }p=1,2,...\leq (n+1)\om+1).
\ee
At these values of the rapidity the S-matrix projects onto the module
of the singlet representation and therefore the poles correspond to
breather states with masses
\[
m_{B^{(a)}_p} = 2 M_a \sin\left(\frac{p\pi}{(n+1)\la}\right)\;,
\]
in which $M_a$ are the quantum masses of the fundamental solitons,
given in (\ref{qumass}).

The pole corresponding to the lowest
breather (the bound state with lowest mass) is $\mu =
(n+1)\om$, which corresponds to $\th =
i\pi(1-\frac2{(n+1)\la})$. Therefore we are able to obtain the S-matrix
element for the scattering of a lowest breather of species $b$ with a
fundamental soliton of species $a$ via the following bootstrap
equation:
\[
S_{A^{(a)}B_1^{(b)}}(\th)\times I_a = I_b\otimes
S_{a,b}(\th+i\pi(\frac12-\frac1{(n+1)\la}))
\,.\, S_{a,b}(\th-i\pi(\frac12-\frac1{(n+1)\la}))\otimes I_b\;.
\]
Thus we need the following formula, derived in appendix
C:
\bea
F_{a,b}(\mu+\frac{n+1}2\om)&\times&F_{a,b}(\mu-\frac{n+1}2\om)= \nn \\
&=& \prod_{k=1}^b \frac {\sin(\pi(\mu+\fo(2k-a-b-1+n)))
\sin(\pi(\mu+\fo(2k-a-b-1-n)))}
{\sin(\pi(\mu+\fo(2k+a-b-1-n)))
\sin(\pi(\mu+\fo(2k+a-b-1+n)))} \nn \\
&&\times \bl \fo(2k+a-b-1-n)\br  \bl
\fo(2k-a-b-3-n)-1 \br \nn \\
&&\times
\bl \fo(2k+a-b-1+n)\br \bl
\fo(2k-a-b+1+n)+1\br\;, \label{fabfab}
\eea
in which we have used the notation
\be
\bl y \br \equiv \frac{\sin(\frac{\pi}{(n+1)\la}(\mu+y))}
{\sin(\frac{\pi}{(n+1)\la}(\mu-y))}\;.
\ee
Combining this with (\ref{sol-br},\ref{smatrix}) we
obtain the S-matrix element
$S_{A^{(a)}B^{(b)}_1}$ for the scattering of an elementary soliton of
species $a$ with a breather $B^{(b)}_1$:
\bea
S_{A^{(a)}B^{(b)}_1}(\th)&=& \prod_{k=1}^b
\bl \fo(2k+a-b-1-n)\br  \bl
\fo(2k-a-b-3-n)-1 \br \nn \\
&&\times
\bl \fo(2k+a-b-1+n)\br \bl
\fo(2k-a-b+1+n)+1\br\;.
\eea
This can be written in a compact
form by introducing the crossing symmetric blocks
\be
\cbl y \cbr \equiv \bl y\br \bl n\om+\om+1-y\br \bl y+n\om\br \bl
w+1-y\br \;,
\label{blocks}
\ee
which have the following properties. They are crossing symmetric
\[
\cbl y\cbr_{\th\to i\pi-\th} = \cbl (n+1)\om+1-y\cbr = \cbl y\cbr
\label{blcrsy}
\]
and $2\pi i$ periodic
\[
\cbl y\cbr_{\th\to\th+2\pi i} = \cbl y+2(n+1)\om+2\cbr = \cbl
y\cbr\;. \label{blperiod}
\]
Therefore we can write
\be
S_{A^{(a)}B^{(b)}_1}(\th)= \prod_{k=1}^b \cbl
\fo(a+b-2k+1-n)\cbr\;. \label{AB1}
\ee
This S-matrix element is just a scalar function which gives the
scattering amplitude for the scattering of a fundamental soliton with
a breather. This can formally be written as a braiding relation:
\be
A^{(a),j}(\th_1)B_1^{(b)}(\th_2) =
S_{A^{(a)}B_1^{(b)}}(\th_1-\th_2)B_1^{(b)}(\th_2)A^{(a),j}(\th_1),
\ee
in which the superscript $j$ denotes the $j$th soliton in the
$a$th multiplet. Since we will make no further use of relations of
this kind, we will always omit the superscript $j$. (For further
details on this Zamolodchikov algebra see for instance \cite{fring95}
or \cite{gande95}).

{}From (\ref{AB1}) we obtain the breather-breather S-matrix by
applying the same fusion procedure again:
\bea
S_{B^{(a)}_1 B^{(b)}_1}(\th) &=&
S_{A^{(a)}B^{(b)}_1}(\mu+\frac{n+1}2\om)
S_{A^{(a)}B^{(b)}_1}(\mu-\frac{n+1}2\om) = \nn \\
&=& \prod_{k=1}^b  \bl \fo(2k-2+a-b)\br
\bl\fo(2k+a-b)\br \nn \\
&&\times \bl\fo(2n+2k-a-b)+1\br
\bl\fo(2n+2k+2-a-b)+1\br \nn \\
&& \times \bl\fo(2k-a-b)+1\br \bl\fo(2k-2-a-b)-1\br
\nn \\
&& \times \bl\fo(2n+2k+2+a-b)+2\br
\bl\fo(2n+2k+a-b)\br= \nn \\
&=& \prod_{k=1}^b \cbl \fo(a+b-2k-2n)\cbr  \cbl
\fo(a+b-2k+2)\cbr\;. \label{sb1b1}
\eea

Now we want to compare this expression with the S-matrix for the
fundamental quantum particles, which was found in \cite{deliu92} for
the real $d_{n+1}^{(2)}$ ATFTs:
\be
S^{(r)}_{ab}(\th) = \prod_{k=1}^b \cbl2k+a-b-1\cbr_H
\cbl H-2k-a+b+1\cbr_H \label{realsmatrix}
\ee
in which
\[
\cbl y\cbr_H = \frac{\bigl( y-1 \bigr)_H \bigl(
y+1\bigr)_H} {\bigl( y-1+B\bigr)_H
\bigl( y+1-B\bigr)_H} \hspace{.7cm}\mbox{and}\hspace{.7cm} \bigl(
y\bigr)_H = \frac{\sin(\frac{\th}{2i}+
\frac{y\pi}{2H})}{\sin(\frac{\th}{2i}- \frac{y\pi}{2H})}\;.
\]
H is twice the quantum Coxeter number, $H=2n+2-2B$ and we assume
that the coupling constant dependent function B is given by $B(\beta) =
\frac{\beta^2}{2\pi+\beta^2}$  ({\em cf} \cite{deliu92} and
\cite{dorey93}, with appropriate normalisations). After analytic
continuation ($\beta \to i\beta$) we are able to make the following
identifications\footnote{For the
S-matrix elements involving higher breathers we were not able to find
a generalisation of the blocks $\{ y\}_H$ used in real affine
Toda theory, and it is for this reason that our definition of the
blocks $\{y\}$ is not related to the blocks $\{ y\}_H$.}:
\bea
H \to (n+1)\frac{\la}{\om},&& \hspace{1cm} B\to -\frac1{\om}, \nn \\
\bl y\br_H &\to& \bl \fo y\br\;. \nn
\eea
Applying this in (\ref{realsmatrix}) it finally emerges that:
\be
S^{(r)}_{ab}(\th) \to
S_{B_1^{(a)}B_1^{(b)}}(\th)\;. \label{identification}
\ee
Thus $S_{B^{(a)}_1 B^{(b)}_1}(\th)$ is indeed identical to
$S^{(r)}_{ab}(\th)$ after analytic continuation ($\beta\to i\beta$)
and we can therefore identify the lowest breather states $B^{(a)}_1$
with the $a$th fundamental quantum
particle of the theory. This generalises results found for
sine-Gordon and $a_2^{(1)}$-ATFT (\cite{gande95}). Since there remains
little doubt that the S-matrix for real ATFT found in \cite{deliu92} is
correct this exact agreement with the lowest breather S-matrix
provides the best possible justification so far for our S-matrix
conjecture.

For completeness we also give the S-matrix elements involving higher
breathers. These were constructed in the same way as shown for
$B_1^{(a)}$ and can all be written conveniently in terms of the blocks
(\ref{blocks}). We obtain the S-matrix elements for the scattering of
a fundamental soliton $A^{(a)}$ with a breather $B_p^{(b)}$:
\be
S_{A^{(a)}B^{(b)}_p}(\th) = \prod_{l=1}^p  \prod_{k=1}^b
\cbl\fo(a+b-2k+1-n)-l+\frac12+\frac p2\cbr\;,
\ee
of two breathers $B_r^{(a)}$ and $B_p^{(b)}$
\be
S_{B^{(a)}_r B^{(b)}_p}(\th) =
\prod_{l=1}^p  \prod_{k=1}^b  \cbl\fo(a+b-2k-2n)-l+\frac{p+r}2\cbr
\cbl\fo(a+b-2k+2)+l-\frac{p+r}2\cbr \label{brbrsmatrix}
\ee
and of an excited soliton $A_r^{(2a)}$ (which will be defined in
the next section) with a breather $B_p^{(b)}$:
\be
S_{A_r^{(2a)}B_p^{(b)}}(\th) = \prod_{l=1}^p \prod_{k=1}^b
\cbl\fo(b-2k-n+1)-l+\frac12+\frac{p+r}2\cbr
\cbl\fo(2a+b-2k-n+1)+l-\frac12-\frac{p+r}2\cbr\;.
\ee

The same fusion procedure could also be applied in order to construct
the S-matrix for the scattering of two excited solitons. This S-matrix
would not be just a scalar function, but, like $S^{a,b}(\th)$, an
intertwiner on the tensor product of the two corresponding
modules. This construction, however, remains beyond the scope of this
paper.
\vspace{2cm}

\section{Pole structure}
In this section we discuss the pole structure of our conjectured
S-matrix.
For the S-matrix to be consistent with the bootstrap
equations we would need to explain its entire pole structure and show
that the bootstrap closes on it. We show explicitly for the case
of $d_3^{(2)}$ ATFT how all poles in the soliton S-matrix can be
explained by fusion into bound states or by higher order diagrams.
We also give some examples of how to explain poles in the bound
state S-matrices. We will see that all important properties of the
general theory already appear in the $d_3^{(2)}$ case and this will
therefore lead us to conjecture the full spectrum of solitons and
bound states and a complete set of three-point couplings in
$d_{n+1}^{(2)}$ ATFT.

\subsection{Example: $d_3^{(2)}$ ATFT}

Before we study the pole structure of the general case in the next
section, we provide here for the sake of clarity a detailed account of
all the poles
in the S-matrices for $d_{3}^{(2)}$.
The notations defined in (\ref{xqdef}) reduce in the case of $n=2$ to
\be
\la = \frac{4\pi}{\beta^2}-\frac{4}{3} \hspace{1cm},\hspace{1cm}
\om = \frac{2\pi}{\beta^2}-1
\ee
and thus $2\om = \la - \frac23$.

The S-matrix elements for the scattering of fundamental solitons can
be written as
\bea
S_{1,1}(\th) &=& F_{1,1}(\mu) \,\tau\{ \check{P}_{2\la_1}+ \lag
2\rag\check{P}_{\la_2} + \lag 6\rag \check{P}_0 \}\tau^{-1} \nn \\
S_{2,1}(\th) &=& -F_{2,1}(\mu) \lag 1\rag \,\tau\{
\check{P}_{\la_1+\la_2} + \lag 5\rag\check{P}_{\la_1} \}\tau^{-1} \nn
\\
S_{2,2}(\th) &=& F_{2,2}(\mu) \lag 0\rag \lag 2\rag \,\tau\{
\check{P}_{2\la_2}+ \lag 4\rag\check{P}_{2\la_1} + \lag 4\rag \lag
6\rag \check{P}_0 \}\tau^{-1}\;
\eea
(where we omit the indices from $\tau$).
Using the expression (\ref{Fab}) one can write down explicit
expressions for the scalar factors in terms of gamma functions. After
a careful study of their pole structure and taking all pole-zero
cancellations into account we obtain the following simple poles on the
physical strip. (For the sake of simplicity we will write the poles in
terms of $\mu = -i\frac{3\la}{2\pi}\th$, such that the physical strip
corresponds to $0 \leq \mbox{Re}\mu \leq \frac32\la = 3\om +1$ and
$\mbox{Im}\mu = 0$.) \vspace{.3cm} \\
In $S_{1,1}(\th)$:
\bea
\mu &=& \om-p \hspace{1.9cm} (\mbox{for }p = 0,1,2,...\leq \om) \nn \\
\mu &=& 3\om+1-p \hspace{1cm} (\mbox{for } p = 1,2,...\leq 3\om +1)
\eea
and their cross channel poles:
\bea
\mu &=& 2\om+1+p \hspace{1cm} (\mbox{for }p = 0,1,2,...\leq \om) \nn \\
\mu &=& p \hspace{2.7cm} (\mbox{for } p = 1,2,...\leq 3\om +1).\nn
\eea
\vspace{.1cm} \\
In $S_{2,1}(\th)$:
\be
\mu = \frac52\om+1-p \hspace{1cm} (\mbox{for }p = 0,1,2,...\leq
\frac52\om+1)
\ee
and their cross channel poles:
\[
\mu = \frac12\om+p \hspace{1.8cm} (\mbox{for }p = 0,1,2,...\leq
\frac52\om+1).
\]
\vspace{.1cm} \\
In $S_{2,2}(\th)$:
\bea
\mu &=& 2\om+1-p \hspace{1cm} (\mbox{for }p = 0,1,2,...\leq 2\om+1) \nn \\
\mu &=& 3\om+1-p \hspace{1cm} (\mbox{for } p = 1,2,...\leq 3\om +1)
\eea
and their cross channel poles:
\bea
\mu &=& \om+p \hspace{1cm} (\mbox{for }p = 0,1,2,...\leq 2\om+1) \nn \\
\mu &=& p \hspace{1.8cm} (\mbox{for } p = 1,2,...\leq 3\om +1).\nn
\eea

Now we will try to explain all these poles either by fusion
processes or higher order scattering diagrams. As already mentioned in
the previous section the poles $\mu=3\om+1-p$
in $S_{1,1}(\th)$ and in $S_{2,2}(\th)$ appear in front of the
projectors onto
the singlet representations and are therefore expected to correspond
to scalar bound states $B_p^{(1)}$ and $B_p^{(2)}$ (for $p=1,2,...\leq
3\om+1$) with masses determined by the mass formula (\ref{massform}).

The second set of poles in $S_{1,1}(\th)$ can be identified as the poles
corresponding to bound states transforming under the module $V_2$. We
will call these bound states excited solitons of type $2$ and denote
them as $A_p^{(2)}(\th)$. The term `excited soliton' was chosen in order
to highlight the fact that the state with $p=0$ is indeed just the
fundamental soliton $A^{(2)}$.

In the light of this last definition one is tempted to identify the poles
in $S_{2,1}$ as corresponding to some sort of excited solitons
transforming under $V_1$, in particular since the lowest pole
$\mu=\frac52\om+1$ indeed corresponds to the fusion
$A^{(1)}+A^{(2)}\to A^{(1)}$. However, closer examination reveals that
for $p\geq 1$ these poles cannot be explained by fusion into bound
states. They do not take part in the bootstrap and one has to find a
different interpretation. We will find that they can be explained by a
subtle generalisation of a mechanism first discovered in the
sine-Gordon theory by Coleman and Thun in \cite{colem78}. The simplest
example for such a generalised Coleman-Thun mechanism
is a crossed box diagram, in which the scattering element in the
middle displays a zero such that the expected double pole from the
diagram is reduced to a simple pole.
This mechanism was first described in
\cite{corri93} in connection with the S-matrices of nonsimply
laced real ATFTs and has also been applied in \cite{hollo94}
and in imaginary $a_2^{(1)}$ ATFT \cite{gande95} to explain several
simple poles in the S-matrix elements.
Figure 1 shows such a crossed box diagram which indeed
corresponds exactly to those poles in $S_{2,1}(\th)$ which we are
trying to explain. If all particles in a diagram like \mbox{figure 1}
are on shell then
the angles (which correspond to purely imaginary rapidity differences)
in the diagram are fixed uniquely and can be calculated by
using elementary geometrical considerations. In figure 1 we obtain
the rapidity difference of the incoming particles to be $\mu
=\fo+p$ in which $p=1,2,...\leq \frac52 \om+1$. Since we also
find that the internal scattering process occurs at
$\mu=\frac52\om+\frac{p}2+\frac12$, at which $S_{A^{(1)}B^{(1)}_p}$
displays a simple zero, we have shown that the diagram in figure 1
explains all remaining poles in $S_{1,2}(\th)$.

\vspace{1cm}
\thinlines
%
%
\begin{center}
\begin{picture}(140,160)(-10,-30)
\put(0,0){\line(1,2){20}}
\put(120,0){\line(-1,2){20}}
\put(20,40){\line(0,1){60}}
\put(20,40){\line(4,3){80}}
\put(20,100){\line(4,-3){80}}
\put(100,40){\line(0,1){60}}
\put(20,100){\line(-1,2){20}}
\put(100,100){\line(1,2){20}}
\put(60,70){\circle*{8}}
\put(-8,-10){\shs{\fns{$A^{(2)}$}}}
\put(115,-10){\shs{\fns{$A^{(1)}$}}}
\put(115,142){\shs{\fns{$A^{(2)}$}}}
\put(-5,142){\shs{\fns{$A^{(1)}$}}}
\put(2,65){\shs{\tiny{$A^{(1)}$}}}
\put(101,65){\shs{\tiny{$A^{(1)}$}}}
\put(35,46){\shs{\tiny{$A^{(1)}$}}}
\put(65,46){\shs{\tiny{$B^{(1)}_p$}}}
\put(36,89){\shs{\tiny{$B^{(1)}_p$}}}
\put(71,89){\shs{\tiny{$A^{(1)}$}}}
\put(30,-30){\shs{\em Figure 1}}
\end{picture}

(In this and all following diagrams time is moving upwards.)
\end{center}
The only set of poles not explained so far are the poles $\mu =
2w+1-p$ in $S_{2,2}(\th)$. The lowest one of these poles ($p=0$)
is explained by another crossed box diagram, this time
with $A^{(2)}$ on all the external and $A^{(1)}$ on all the
internal legs (see figure 2).
Here we encounter a slight variation of the generalised Coleman-Thun
mechanism first pointed out by Hollowood \cite{hollo94}.
This diagram again gives a na\"{\i}ve double pole, and here
the internal $S$-matrix element is non-zero. That the diagram
nevertheless corresponds to a simple pole is owing to the fact that,
although non-zero, the internal $S_{1,1}$ projects onto $V_2$,
which is not present in the external legs' $V_2\otimes V_2$
because of truncation of the TPG. This mechanism generalises
to all the non-particle $p=0$ poles and is explained in appendix
\ref{poles}, where it is also used to explain the rational
$S$-matrix pole structure.

\vspace{1cm}
%
\begin{center}
\begin{picture}(140,160)(-10,-30)
\put(0,0){\line(1,2){20}}
\put(120,0){\line(-1,2){20}}
\put(20,40){\line(0,1){60}}
\put(20,40){\line(4,3){80}}
\put(20,100){\line(4,-3){80}}
\put(100,40){\line(0,1){60}}
\put(20,100){\line(-1,2){20}}
\put(100,100){\line(1,2){20}}
\put(60,70){\circle*{8}}
\put(-8,-10){\shs{\fns{$A^{(2)}$}}}
\put(115,-10){\shs{\fns{$A^{(2)}$}}}
\put(115,142){\shs{\fns{$A^{(2)}$}}}
\put(-5,142){\shs{\fns{$A^{(2)}$}}}
\put(4,65){\shs{\tiny{$A^{(1)}$}}}
\put(101,65){\shs{\tiny{$A^{(1)}$}}}
\put(35,46){\shs{\tiny{$A^{(1)}$}}}
\put(65,46){\shs{\tiny{$A^{(1)}$}}}
\put(36,89){\shs{\tiny{$A^{(1)}$}}}
\put(71,89){\shs{\tiny{$A^{(1)}$}}}
\put(30,-30){\shs{\em Figure 2}}
\end{picture}
\end{center}
The rest of the poles (for $p>0$), however,
correspond to an even more complicated generalised Coleman-Thun
mechanism depicted in figure 3. In this diagram we find the rapidity
difference of the two incoming particles $A^{(2)}$ to be
$\mu=2\om+1-p$ in which $p$ can take values $1,2,...\leq
2\om+1$. These are exactly the poles in $S_{2,2}(\th)$ which we are
trying to explain. However, the diagram in figure 3 contains five
loops and thirteen
internal lines and should therefore lead to cubic poles. The black
dots in the diagram represent scattering processes of the internal
particles. Calculating the internal angles of the diagram it emerges
that the internal $A^{(1)}-A^{(1)}$ scattering process occurs at the
rapidity difference $\mu = \om+1-p$ at which $S_{1,1}(\th)$ has neither
pole nor zero. The two $B^{(1)}_p-A^{(1)}$ processes, however, occur at
a rapidity difference of $\mu=\fo+\frac12-\frac{p}2$ and from
(\ref{AB1}) we can see that at exactly those values
$S_{A^{(1)}B^{(1)}_p}(\th)$ displays a simple zero, which reduces the
expected cubic pole to the observed simple poles in $S_{2,2}(\th)$.

\vspace{1cm}
%
\begin{center}
\begin{picture}(220,170)(-10,-30)
\put(0,0){\line(3,2){24}}
\put(24,16){\line(1,1){128}}
\put(24,16){\line(4,1){88}}
\put(112,37){\line(0,1){86}}
\put(152,16){\line(3,-2){24}}
\put(112,37){\line(2,-1){40}}
\put(0,160){\line(3,-2){24}}
\put(24,144){\line(1,-1){128}}
\put(24,144){\line(4,-1){88}}
\put(112,123){\line(2,1){40}}
\put(152,144){\line(3,2){24}}
\put(88,80){\circle*{8}}
\put(112,56){\circle*{8}}
\put(112,104){\circle*{8}}
\put(-6,-10){\shs{\fns{$A^{(2)}$}}}
\put(173,-10){\shs{\fns{$A^{(2)}$}}}
\put(-5,163){\shs{\fns{$A^{(2)}$}}}
\put(175,163){\shs{\fns{$A^{(2)}$}}}
\put(45,52){\shs{\tiny{$A^{(1)}$}}}
\put(43,104){\shs{\tiny{$A^{(1)}$}}}
\put(70,21){\shs{\tiny{$A^{(1)}$}}}
\put(120,19){\shs{\tiny{$A^{(1)}$}}}
\put(127,42){\shs{\tiny{$A^{(1)}$}}}
\put(113,77){\shs{\tiny{$B^{(1)}_p$}}}
\put(128,116){\shs{\tiny{$A^{(1)}$}}}
\put(70,134){\shs{\tiny{$A^{(1)}$}}}
\put(123,134){\shs{\tiny{$A^{(1)}$}}}
\put(60,-30){\shs{\em Figure 3}}
\end{picture}
\end{center}

The pole structure of the S-matrix lead us to conjecture the
following particle spectrum of $d_3^{(2)}$ ATFT:
\vspace{.3cm} \\
1) {\em fundamental solitons} $A^{(a)}$ $(a=1,2)$: \\
 \hspace*{1cm} masses $M_a = C8\sqrt{2}\frac{3m}{\beta^2}
\sin(\frac{a\pi}3(\frac12-\frac1{3\la}))$ \vspace{0.3cm} \\
2) {\em breathers} $B^{(a)}_p$ ($A^{(a)}-A^{(a)}$ bound states)
($a=1,2$ and $p = 1,2,...\leq 3\om+1$):\\
 \hspace*{1cm}masses $m_{B_p^{(a)}} = 2M_a\sin(\frac{p\pi}{3\la})$
\vspace{0.3cm} \\
3) {\em breathing solitons} $A^{(2)}_p$ ($A^{(1)}-A^{(1)}$ bound
states) ($p = 0,1,2,...\leq \om$)\\
\hspace*{1cm} masses $m_{A^{(2)}_p} =
2M_1\cos(\frac{\pi}6-\frac{\pi}{9\la}-\frac{\pi p}{3\la})$.
\vspace{.5cm} \\
Here we should mention a problem that occured in
\cite{hollo93b}. Hollowood rejected his $c_n$ invariant S-matrix
because there was a pole at $\th=i\frac{\pi}2$ in $S_{a,n+1-a}$ which
could not be explained in terms of the particle spectrum of the
theory. Here we can see that in the weak coupling limit
($\la\to\infty$) the element $S_{2,1}$ has poles only at
$\th=i\frac{5\pi}6$ and $i\frac{\pi}6$. So due to the definitions
(\ref{xqdef}) and the fact that we do not need to include an
additional minimal Toda factor, the poles, which caused serious
problems in \cite{hollo94}, do not appear in our S-matrix conjecture.

We expect that it will prove possible to explain all poles in the
S-matrix elements involving breathers in a similar way. We will show
this for one example, the scattering of the two breathers $B_r^{(1)}$
and $B_p^{(1)}$. Without loss of generality we choose $r\geq p$. The
S-matrix element $S_{B_r^{(1)}B_p^{(1)}}(\th)$ was given in
(\ref{brbrsmatrix}) and in the case of $n=2$ contains the following
simple and double poles on the physical strip:
\bea
\mu^{(1)}_l &=& \om-l+1-\frac{r-p}2 \nn \\
\mu^{(2)}_l &=& 2\om+l+\frac{r-p}2 \hspace{1cm} (\mbox{simple poles for
}l=1,2,...,p) \nn
\eea
and
\bea
\mu^{(3)}_l &=& 3\om+l+1-\frac{r+p}2 \hspace{1cm} \nn \\
\mu^{(4)}_l &=& -l+\frac{r+p}2 \hspace{1.5cm}(\mbox{double poles if }
l=1,2,...,p-1; \nn \\
&&\hspace{4cm} \mbox{simple poles if } l=0,p)\;. \nn
\eea
By using the mass formula (\ref{massform}) we obtain that two
breathers can fuse together to build another breather only if either
the species or the excitation numbers of the two incoming breathers
are identical. Thus in general the only possible fusion processes of
two breathers are $B_r^{(a)} + B_p^{(a)} \to B_{r\pm p}^{(a)}$ and
$B_p^{(a)} + B_p^{(b)} \to B_p^{(a\pm b)}$ (see section 4.2, figure 5d
and 5e).
Therefore in our example we have two possible fusion processes, i.e.
$B_r^{(1)} + B_p^{(1)} \to B_{r\pm p}^{(1)}$, which explain the simple
poles $\mu_l^{(3)}$ and $\mu_l^{(4)}$ for $l= 0,p$.

The double poles $\mu_l^{(3)}$ and $\mu_l^{(4)}$ are the direct and
cross channel poles corresponding to the crossed box diagram in figure
4a. Unlike in the case of figure 1, the process in the centre of the
diagram occurs at a rapidity difference where the corresponding
S-matrix has neither pole nor zero, and therefore the process in
figure 4a leads to a double pole.

In order to explain the remaining poles $\mu_l^{(1)}$ and
$\mu_l^{(2)}$ we find another third order diagram, depicted in figure
4b (in which $s$ can take values $s = 1,2,...,p$). Here again two of
the three internal scattering processes display a simple zero, which
reduces the expected triple poles to simple poles.

\vspace{1cm}
%
\begin{center}
\begin{picture}(420,170)(-10,-30)
\put(0,0){\line(1,2){20}}
\put(120,0){\line(-1,2){20}}
\put(20,40){\line(0,1){60}}
\put(20,40){\line(4,3){80}}
\put(20,100){\line(4,-3){80}}
\put(100,40){\line(0,1){60}}
\put(20,100){\line(-1,2){20}}
\put(100,100){\line(1,2){20}}
\put(60,70){\circle*{8}}
\put(-8,-10){\shs{\fns{$B_r^{(1)}$}}}
\put(115,-10){\shs{\fns{$B_p^{(1)}$}}}
\put(115,142){\shs{\fns{$B_r^{(1)}$}}}
\put(-5,142){\shs{\fns{$B_p^{(1)}$}}}
\put(4,65){\shs{\tiny{$B_s^{(1)}$}}}
\put(102,65){\shs{\tiny{$B_s^{(1)}$}}}
\put(35,46){\shs{\tiny{$B_{r-s}^{(1)}$}}}
\put(66,46){\shs{\tiny{$B_{p-s}^{(1)}$}}}
\put(34,91){\shs{\tiny{$B_{p-s}^{(1)}$}}}
\put(66,91){\shs{\tiny{$B_{r-s}^{(1)}$}}}
\put(30,-30){\shs{\em Figure 4a}}
%
%
%
\put(200,0){\line(3,2){24}}
\put(224,16){\line(1,1){128}}
\put(224,16){\line(4,1){88}}
\put(312,37){\line(0,1){86}}
\put(352,16){\line(3,-2){24}}
\put(312,37){\line(2,-1){40}}
\put(200,160){\line(3,-2){24}}
\put(224,144){\line(1,-1){128}}
\put(224,144){\line(4,-1){88}}
\put(312,123){\line(2,1){40}}
\put(352,144){\line(3,2){24}}
\put(288,80){\circle*{8}}
\put(312,56){\circle*{8}}
\put(312,104){\circle*{8}}
\put(194,-10){\shs{\fns{$B_r^{(1)}$}}}
\put(373,-10){\shs{\fns{$B_p^{(1)}$}}}
\put(195,163){\shs{\fns{$B_p^{(1)}$}}}
\put(375,163){\shs{\fns{$B_r^{(1)}$}}}
\put(238,52){\shs{\tiny{$B_{r-s}^{(1)}$}}}
\put(242,104){\shs{\tiny{$B_{p-s}^{(1)}$}}}
\put(270,20){\shs{\tiny{$B_s^{(1)}$}}}
\put(320,18){\shs{\tiny{$B_s^{(1)}$}}}
\put(327,42){\shs{\tiny{$B_{p-s}^{(1)}$}}}
\put(313,77){\shs{\tiny{$B_s^{(2)}$}}}
\put(328,116){\shs{\tiny{$B_{r-s}^{(1)}$}}}
\put(270,134){\shs{\tiny{$B_s^{(1)}$}}}
\put(323,136){\shs{\tiny{$B_s^{(1)}$}}}
\put(265,-30){\shs{\em Figure 4b}}
\end{picture}
\end{center}
In the following section we will extend this discussion to the general case
of $d_{n+1}^{(2)}$ ATFT.
\subsection{$d_{n+1}^{(2)}$ ATFT}
{}From our observations in the $d_3^{(2)}$ case we are able to
conjecture the following particle spectrum for imaginary
$d_{n+1}^{(2)}$ ATFT: \vspace{.1cm} \\
1) {\em fundamental solitons} $A^{(a)}$ $(a=1,2,..,n)$: \\
 \hspace*{1cm} masses $M_a = C8\sqrt{2}\frac{(n+1)m}{\beta^2}
\sin(\frac{a\pi}{n+1}(\frac12-\frac1{(n+1)\la}))$ \vspace{0.1cm} \\
2) {\em breathers} $B^{(a)}_p$ ($A^{(a)}-A^{(a)}$ bound states)
($a=1,2,...,n$ and $p = 1,2,...\leq (n+1)\om+1$):\\
 \hspace*{1cm}masses $m_{B_p^{(a)}} = 2M_a\sin(\frac{p\pi}{(n+1)\la})$
\vspace{0.1cm} \\
3) {\em breathing solitons} $A^{(2a)}_p$ ($A^{(a)}-A^{(a)}$ bound
states) ($p = 0,1,2,...\leq a\om$)\\
\hspace*{1cm} masses $m_{A^{(2a)}_p} =
2M_a\cos(\frac{a\pi}{n+1}(\frac12-\frac1{(n+1)\la}-\frac{p}{a\la}))$.

We have not examined the full pole structure of all S-matrix
elements explicitly. Generalising the results of the $d_3^{(2)}$
case, however, we conjecture a list of all possible three point
vertices in the theory. These six
vertices seem to be the only vertices which are consistent with the
mass formula (\ref{massform}) and the pole structure of the S-matrix.
Although we are not able to give a general proof that this list is
complete, all poles and diagrams we were able to examine
could be explained by using only these vertices. (In the
diagrams we have used the abbreviation $\tm \equiv
\mu(i\pi)=(n+1)\om+1$):

\vspace{1cm}
%
%
%
\begin{center}
\begin{picture}(420,120)(-10,-30)
%
%
\put(0,0){\line(1,1){40}}
\put(40,40){\line(1,-1){40}}
\put(40,40){\line(0,1){64}}
\put(40,40){\circle{16}}
\put(40,17){\vector(0,1){12}}
\put(22,58){\vector(1,-1){10}}
\put(58,58){\vector(-1,-1){10}}
\put(30,7){\shs{\scs{$\frac{a+b}2\om$}}}
\put(0,64){\shs{\scs{$\tm-\frac{b}2\om$}}}
\put(53,64){\shs{\scs{$\tm-\frac{a}2\om$}}}
\put(-5,-10){\shs{\fns{$A^{(a)}$}}}
\put(76,-10){\shs{\fns{$A^{(b)}$}}}
\put(35,108){\shs{\fns{$A^{(a+b)}$}}}
%
%
%
\put(160,0){\line(1,1){40}}
\put(200,40){\line(1,-1){40}}
\put(200,40){\line(0,1){64}}
\put(200,40){\circle{16}}
\put(200,17){\vector(0,1){12}}
\put(182,58){\vector(1,-1){10}}
\put(218,58){\vector(-1,-1){10}}
\put(186,7){\shs{\scs{$\tm-p$}}}
\put(161,64){\shs{\scs{$\frac{\tm}2+\frac{p}2$}}}
\put(214,64){\shs{\scs{$\frac{\tm}2+\frac{p}2$}}}
\put(155,-10){\shs{\fns{$A^{(a)}$}}}
\put(236,-10){\shs{\fns{$A^{(a)}$}}}
\put(195,108){\shs{\fns{$B_p^{(a)}$}}}
%
%
\put(320,0){\line(1,1){40}}
\put(360,40){\line(1,-1){40}}
\put(360,40){\line(0,1){64}}
\put(360,40){\circle{16}}
\put(360,17){\vector(0,1){12}}
\put(342,58){\vector(1,-1){10}}
\put(378,58){\vector(-1,-1){10}}
\put(348,7){\shs{\scs{$a\om-p$}}}
\put(302,64){\shs{\scs{$\tm-\frac{a}2\om+\frac{p}2$}}}
\put(369,64){\shs{\scs{$\tm-\frac{a}2\om+\frac{p}2$}}}
\put(315,-10){\shs{\fns{$A^{(a)}$}}}
\put(396,-10){\shs{\fns{$A^{(a)}$}}}
\put(355,108){\shs{\fns{$A^{(2a)}_p$}}}
\put(17,-30){\shs {\em Figure 5a}}
\put(177,-30){\shs {\em Figure 5b}}
\put(337,-30){\shs {\em Figure 5c}}
\end{picture}

\vspace{1.5cm}
\begin{picture}(420,120)(-10,-30)
%
%
\put(0,0){\line(1,1){40}}
\put(40,40){\line(1,-1){40}}
\put(40,40){\line(0,1){64}}
\put(40,40){\circle{16}}
\put(40,17){\vector(0,1){12}}
\put(22,58){\vector(1,-1){10}}
\put(58,58){\vector(-1,-1){10}}
\put(33,5){\shs{\scs{$\frac{r+p}2$}}}
\put(2,64){\shs{\scs{$\tm-\frac{r}2$}}}
\put(53,64){\shs{\scs{$\tm-\frac{p}2$}}}
\put(-5,-10){\shs{\fns{$B^{(a)}_p$}}}
\put(76,-10){\shs{\fns{$B^{(a)}_r$}}}
\put(35,108){\shs{\fns{$B^{(a)}_{p+r}$}}}
%
%
%
\put(160,0){\line(1,1){40}}
\put(200,40){\line(1,-1){40}}
\put(200,40){\line(0,1){64}}
\put(200,40){\circle{16}}
\put(200,17){\vector(0,1){12}}
\put(182,58){\vector(1,-1){10}}
\put(218,58){\vector(-1,-1){10}}
\put(192,5){\shs{\scs{$\frac{a+b}2\om$}}}
\put(156,64){\shs{\scs{$\tm-\frac b2\om$}}}
\put(211,64){\shs{\scs{$\tm-\frac a2\om$}}}
\put(155,-10){\shs{\fns{$B^{(a)}_p$}}}
\put(236,-10){\shs{\fns{$B^{(b)}_p$}}}
\put(195,108){\shs{\fns{$B^{(a+b)}_p$}}}
%
%
\put(320,0){\line(1,1){40}}
\put(360,40){\line(1,-1){40}}
\put(360,40){\line(0,1){64}}
\put(360,40){\circle{16}}
\put(360,17){\vector(0,1){12}}
\put(342,58){\vector(1,-1){10}}
\put(378,58){\vector(-1,-1){10}}
\put(345,7){\shs{\scs{$\tm-\frac{r-p}2$}}}
\put(310,64){\shs{\scs{$\frac{\tm}2-\frac a2\om+\frac r2$}}}
\put(367,64){\shs{\scs{$\frac{\tm}2+\frac a2\om-\frac p2$}}}
\put(315,-10){\shs{\fns{$A_p^{(2a)}$}}}
\put(396,-10){\shs{\fns{$A_r^{(2a)}$}}}
\put(355,108){\shs{\fns{$B_{p-r}^{(a)}$}}}
\put(17,-30){\shs {\em Figure 5d}}
\put(177,-30){\shs {\em Figure 5e}}
\put(337,-30){\shs {\em Figure 5f}}
\end{picture}
\end{center}
Due to the bootstrap principle all fusion processes obtained by
turning any of the above diagrams by $120$ degrees are also
possible. It might seem a surprise that there are only excitations
of solitons of even species (figure 5c). However, in \cite{harde94}
the bound states of the classical soliton solutions of $a_n^{(1)}$
ATFTs were studied and it was found that only solitons of species $2a
\mbox{mod}h$ can be excited. Since the `$\mbox{mod}h$' results from
non-self-conjugacy of the representations of the $a_n$ algebras,
we expect that for $d_{n+1}^{(2)}$ algebras (even in the
classical case) excited solitons exist for even species only.

It should also be noted that, as already discovered in the $a_2^{(1)}$
case \cite{gande95} and for the classical solitons of $a_n^{(1)}$ AFTF
\cite{harde94}, only fundamental solitons of the same species (and of
conjugate species in non-self-conjugate theories) can form bound
states (figures 5b and 5c). We expect this to be a common
feature of all imaginary ATFTs.

\section{Discussion}

In this paper we have demonstrated how to construct consistent soliton
S-matrices
for $d_{n+1}^{(2)}$ affine Toda field theories by using trigonometric
solutions of the Yang-Baxter equation, previously only
carried out successfully for $a_n^{(1)}$ ATFTs. The crux of
the extension to nonsimply-laced algebras is the fact that the
mass ratios of the solitons change under
renormalisation. This is accommodated by changing the dependence
of the spectral parameter $x$
and the deformation parameter $q$ on the physical variables $\beta^2$
and $\th$ in equation (\ref{xqdef}) (see also
\cite{babic94,deliu95,efthi93,smirn91}).
After finding an appropriate
overall scalar factor we constructed the S-matrices for all scattering
processes involving breathers, and we were able to show that the
S-matrix of the lowest breather states coincides with the exact
S-matrix for the fundamental quantum particles
(\ref{identification}). This identification
has so far only been shown for sine-Gordon theory and in the case of
$a_2^{(1)}$.
Because of the self-conjugacy of $d_{n+1}^{(2)}$ ATFT the calculations
in this paper were slightly easier than in the case of $a_n^{(1)}$
theory. However, we have obtained some preliminary results in
demonstrating the particle-breather identification for $a_n^{(1)}$
ATFT which will appear in a forthcoming paper.

\subsection{Duality}

There are two ways in which these theories might exhibit some
form of duality: affine duality, in which the weak regime
of the $\hat{g}$ theory is related to the strong regime
of the $\hat{g}^\vee$ theory, and Lie duality, in which
objects in the $g^{(1)}$ theory are related to others in
the $g^{(1)\vee}$ theory, probably in the same regime.
We can examine the possibilities within the following scheme,
which is intended only as a tentative framework for discussion:
\[
\begin{array}{|ccc|}
\hline
 \;\;{\rm\em weak/strong} & \leftrightarrow & {\rm\em strong/weak}\;\;

\\[0.1in]
g^{(1)} \;\;{\rm solitons} & \leftrightarrow &
g^{\vee (1) \vee} \;\;{\rm solitons} \\[0.1in]
g^{(1)} \;\;{\rm breathers} & \leftrightarrow &
g^{\vee (1) \vee} \;\;{\rm breathers} \\
\updownarrow & & \parallel \\
g^{\vee(1)} \;\;{\rm particles} & \leftrightarrow &
g^{\vee (1) \vee} \;\;{\rm particles} \\[0.1in]

U_q(g^{(1)\vee}) & \supset \;\; U_q(g^\vee) \;\;\subset &
U_q(g^{\vee(1)}) \\[0.1in]
\hline
\end{array}
\]
where $g^{\vee (1) \vee}$ means $((g^\vee)^{(1)})^\vee$ and so on.

In this paper we have examined the right-hand column in the weak
regime and identified the breathers with the particles.
It is unclear whether a strong regime exists: as we
pointed out in appendix \ref{data},
both $\lambda$ and $\omega$ become singular
at finite $\bt$, and breathers and excited solitons disappear
{}from the spectrum. However, there is certainly affine duality
between the particles on the right at coupling $4\pi/i\bt$ and
on the left at $i\bt$ \cite{brade90}. If we formally examine
the soliton and breather masses on the right at $4\pi/i\bt$
we find that the mass ratios are consistent with this scheme
-{\em i.e.\ }are the same for {\em all} the objects -
if both the coupling for solitons and breathers on the left is $\bt$
and, as mentioned before, the soliton mass corrections on
the left are as in the note to \cite{macka94}, a matter which
still remains for us to clarify.

There is, therefore, a possible Lie duality between $(g^\vee,i\bt)$
particles and $(g,\bt)$ breathers, extendable to the original
classical Lie duality \cite{olive93}
suggested between $g$ solitons and $g^\vee$ particles which
would also apply in the quantum case between $(g,\bt)$
solitons and $(g^\vee,i\bt)$ particles \cite{deliu94b}.
For the solitons, a {\em sine qua non} for affine duality is
that the $S$-matrices should have the same algebraic structure.
This is guaranteed by the observation in the last row:
that not only $g^\vee \subset g^{\vee(1)}$ but also
$g^\vee \subset g^{(1)\vee}$. In our example, $g=b_n$
and $g^{\vee(1)\vee}=d_{n+1}^{(2)}$, and the observation
is that, with the appropriate choice of affine root,
$c_n\subset a_{2n-1}^{(2)}$. However, it is then difficult
to see in what sense the two charge algebras could be dual, although
a mechanism might be provided by \cite{deliu95b}.
(Further, if the note in \cite{artz95} is correct and
the $U_q(c_n)$-invariant $R$-matrix which must be used
to describe $b_n^{(1)}$ solitons is obtained from that
for the $d_{n+1}^{(2)}$ solitons by setting $q^{h'}\mapsto -q^{h'-2}$,
it is difficult to see how the correct masses could be obtained.
This is connected with the difficulties described in appendix
\ref{data} and which we hope to resolve in future work.)

Whereas in the particle case affine duality included
identification of the $S$-matrices (possible since
the particles have masses of order $m$ in both weak and strong
regime), this cannot be so simple for the solitons, which,
if they exist in the strong regime, will have light masses
in contrast to their heavy masses (of order $m/\bt^2$) in the
weak regime. As we point out in appendix \ref{data}, this manifests
itself in an overall rescaling of $\la$ and $\om$ in the strong regime
which gives the correct soliton mass ratios but forbids
the existence of excited states, and which makes the leading-order
$R$-matrix structure of the $S$-matrix trivial.
A possibility for incorporating Lie, affine and strong-weak duality
might therefore be that between the $(g^{\vee(1)\vee},4\pi/i\bt)$
solitons and the $(g^{\vee(1)},i\bt)$ particles. This and other possible
relations might be explored by truncating the theory to admit only
soliton degrees of freedom and then examining the relations between
the semiclassical limits of the soliton and particle $S$-matrices
and the soliton time delay \cite{fring94,hollo93,spence95}.
For clarity we illustrate the possibilities as they apply to
our example, $g=b_n$:

\[
\begin{array}{|ccc|}
\hline

b_n^{(1)},\bt \;\;\;{\rm solitons} & \leftrightarrow &
d_{n+1}^{(2)},{4\pi\over i\bt} \;\;\;{\rm solitons} \\[0.1in]
b_n^{(1)},\bt \;\;\;{\rm breathers} & \leftrightarrow &
d_{n+1}^{(2)},{4\pi\over i\bt}\;\;\;{\rm breathers} \\
\updownarrow & & \parallel \\
c_n^{(1)},i\bt \;\;\;{\rm particles} & \leftrightarrow &
d_{n+1}^{(2)},{4\pi\over i\bt} \;\;\;{\rm particles} \\[0.1in]

U_q(a_{2n-1}^{(2)}) & \supset \;\; U_q(c_n) \;\;\subset &
U_q(c_n^{(1)}) \\[0.1in]
\hline
\end{array}
\]

\noindent It will require much more work on the untwisted nonsimply-laced
solitons and their \mbox{$S$-matrices}, however, to place such speculation on
a firmer footing.

\subsection{More open problems}

We do not know a general method of determining which fusion processes
are allowed and so cannot prove that the suggested list
of three-point vertices in section 4.2 is complete.  It would be
interesting to investigate whether there is a
way of obtaining all possible three-point vertices from the properties
of the underlying quantum group symmetry. This could eventually lead
to a generalisation of Dorey's fusing rule \cite{dorey91},
which for real ATFT connects the possible fusion processes with
properties of the root system of the associated Lie algebra.
(For recent work on this fusing rule as it applies to decomposition
of tensor products of representations of quantized affine algebras
and Yangians see \cite{chari95}.) In order to do this we will also
need a better understanding of what conserved quantities
distinguish bound states of the same species. In particular,
it seems important to understand the relationships between the
commuting, local conserved charges of the particles \cite{brade90}
(which have spins equal to the exponents of the algebra), those of
the solitons \cite{free94}, and the non-commuting, non-local
$U_q(\hat{g}^\vee)$ charges.
Finally, on a mathematical level we have no idea how the excitation
level $p$ is related to quantized affine algebra representations.

Another unsolved problem in the imaginary ATFTs is that of unitarity.
The complexity of the Hamiltonian implies a non-unitary theory, but
we nevertheless believe that some truncation to a unitary
theory of the solitons and their bound states will prove possible
\cite{spence95}. The $S$-matrices we have constructed
obey the usual two-particle real unitarity, but we have not
been able to examine
the residues of the poles in sufficient detail to investigate
one-particle unitarity (a requirement if the bootstrap
principle is to be implemented properly). An investigation of the
residues of poles would also be necessary in order to obtain a
general criterion for which
simple poles take part in the bootstrap and which are
explained by some generalised Coleman-Thun mechanism, a connection
found for real ATFTs in \cite{corri93}.

As mentioned previously, it would also be interesting to check the
consistency of our S-matrix conjecture with semiclassical
calculations, as done for $a^{(1)}_n$ in \cite{hollo93}.
The extension of this in particular to
untwisted theories might help us to clarify the semiclassical
mass calculations of \cite{deliu94} and \cite{macka94}.

\vspace{0.3in}
{\bf Acknowledgments}

GMG would like to thank G. Delius for discussions. NJM would like to
thank G. Delius, T. Hollowood and G. Watts for discussions, M. Jimbo
for a helpful communication and the UK PPARC for a research fellowship.

\newpage
%
%
\appendix
\section{$S$-matrix parameters for general algebras}\label{data}

{
\renewcommand{\arraystretch}{1.5}
\headsep -0.5in

\begin{table}
\[
\begin{array}{|ccc|cc|cc|c|}
\hline
 g & & g^\vee & h & h^\vee & h' & h'^\vee
& t \\ \hline \hline

a_n^{(1)} & \begin{array}{c}
\begin{picture}(100,35)(0,0)
\put(55,23){\circle{3}}
\put(10,8){\circle{3}}
\put(25,8){\circle{3}}
\put(40,8){\circle{3}}
\put(70,8){\circle{3}}
\put(85,8){\circle{3}}
\put(100,8){\circle{3}}
\put(11.5,8){\line(1,0){12}}
\put(26.5,8){\line(1,0){12}}
\put(71.5,8){\line(1,0){12}}
\put(86.5,8){\line(1,0){12}}
\put(12,9){\line(3,1){41}}
\put(98,9){\line(-3,1){41}}
\put(48,8){$\dots$}
\put(31,25){$\ss0$}
\put(7,-3){$\ss1$}
\put(22,-3){$\ss2$}
\put(37,-3){$\ss3$}
\put(100,-3){$\ss n$}
\end{picture}%
\end{array}%
& a_n^{(1)} & n+1 & n+1 & n+1 & n+1   & 1
\\  \hline

d_n^{(1)} &
\begin{array}{c}
\begin{picture}(110,45)(0,-10)
\put(14,20){\circle{3}}
\put(14,-4){\circle{3}}
\put(25,8){\circle{3}}
\put(40,8){\circle{3}}
\put(70,8){\circle{3}}
\put(85,8){\circle{3}}
\put(97,19){\circle{3}}
\put(97,-4){\circle{3}}

\put(15,19){\line(1,-1){10}}
\put(15,-3){\line(1,1){10}}
\put(26.5,8){\line(1,0){12}}
\put(71.5,8){\line(1,0){12}}
\put(86,9){\line(1,1){10}}
\put(86,7){\line(1,-1){10}}

\put(48,8){$\dots$}
\put(5,-6){$\ss0$}
\put(5,17){$\ss1$}
\put(23,-3){$\ss2$}
\put(37,-3){$\ss3$}
\put(102,17){$\ss n-1$}
\put(102,-6){$\ss n$}
\end{picture}
\end{array}%
& d_n^{(1)} & 2n-2 & 2n-2 & 2n-2 & 2n-2   & 1
\\  \hline

a_{2n}^{(2)} &
\begin{array}{c}
\begin{picture}(110,25)(0,0)
\put(10,8){\circle{3}}
\put(25,8){\circle{3}}
\put(40,8){\circle{3}}
\put(70,8){\circle{3}}
\put(85,8){\circle{3}}
\put(100,8){\circle{3}}
\put(11,7){\line(1,0){13}}
\put(11,9){\line(1,0){13}}
\put(26.5,8){\line(1,0){12}}
\put(71.5,8){\line(1,0){12}}
\put(86,7){\line(1,0){13}}
\put(86,9){\line(1,0){13}}
\put(89,5){$>$}
\put(14,5){$>$}
\put(48,8){$\dots$}
\end{picture}
\end{array}%
& a_{2n}^{(2)} & 2n+1 & 2n+1 & 2n+1 & 2n+1   & 1
\\ \hline

a_{2n-1}^{(2)} &
\begin{array}{c}
\begin{picture}(110,35)(0,-5)
\put(10,8){\circle{3}}
\put(25,8){\circle{3}}
\put(40,8){\circle{3}}
\put(70,8){\circle{3}}
\put(85,8){\circle{3}}
\put(97,19){\circle{3}}
\put(97,-4){\circle{3}}
\put(11,7){\line(1,0){13}}
\put(11,9){\line(1,0){13}}
\put(26.5,8){\line(1,0){12}}
\put(71.5,8){\line(1,0){12}}
\put(86,9){\line(1,1){10}}
\put(86,7){\line(1,-1){10}}
\put(48,8){$\dots$}
\put(15,5){$>$}
\end{picture}
\end{array}%
& b_n^{(1)}  & 2n-1 & 2n &  2n-1 & 2n  & 1
\\ \hline

d_{n+1}^{(2)} &
\begin{array}{c}
\begin{picture}(110,20)(0,0)
\put(10,8){\circle{3}}
\put(25,8){\circle{3}}
\put(40,8){\circle{3}}
\put(70,8){\circle{3}}
\put(85,8){\circle{3}}
\put(100,8){\circle{3}}
\put(11,7){\line(1,0){13}}
\put(11,9){\line(1,0){13}}
\put(26.5,8){\line(1,0){12}}
\put(71.5,8){\line(1,0){12}}
\put(86,7){\line(1,0){13}}
\put(86,9){\line(1,0){13}}
\put(89,5){$>$}
\put(14,5){$<$}
\put(48,8){$\dots$}
\end{picture}
\end{array}%
& c_n^{(1)} & n+1 & 2n & 2n+2 & 2n & 2
\\ \hline

c_n^{(1)} &
\begin{array}{c}
\begin{picture}(110,25)(0,0)
\put(10,8){\circle{3}}
\put(25,8){\circle{3}}
\put(40,8){\circle{3}}
\put(70,8){\circle{3}}
\put(85,8){\circle{3}}
\put(100,8){\circle{3}}
\put(11,7){\line(1,0){13}}
\put(11,9){\line(1,0){13}}
\put(26.5,8){\line(1,0){12}}
\put(71.5,8){\line(1,0){12}}
\put(86,7){\line(1,0){13}}
\put(86,9){\line(1,0){13}}
\put(89,5){$<$}
\put(14,5){$>$}
\put(48,8){$\dots$}
\put(7,-3){$\ss0$}
\put(22,-3){$\ss1$}
\put(37,-3){$\ss2$}
\put(100,-3){$\ss n$}
\end{picture}
\end{array}%
& d_{n+1}^{(2)} & 2n & n+1 & 2n & 2n+2  & {1\over 2}
\\ \hline

b_n^{(1)} &
\begin{array}{c}
\begin{picture}(110,35)(0,-5)
\put(10,8){\circle{3}}
\put(25,8){\circle{3}}
\put(40,8){\circle{3}}
\put(70,8){\circle{3}}
\put(85,8){\circle{3}}
\put(97,19){\circle{3}}
\put(97,-4){\circle{3}}
\put(11,7){\line(1,0){13}}
\put(11,9){\line(1,0){13}}
\put(26.5,8){\line(1,0){12}}
\put(71.5,8){\line(1,0){12}}
\put(86,9){\line(1,1){10}}
\put(86,7){\line(1,-1){10}}
\put(48,8){$\dots$}

\put(15,5){$<$}
\put(102,-6){$\ss0$}
\put(102,17){$\ss1$}
\put(80,-3){$\ss2$}
\put(7,-3){$\ss n$}
\end{picture}
\end{array}%
& a_{2n-1}^{(2)} & 2n & 2n-1 & 2n & 2n-1  & 1
\\ \hline
\end{array}
\]
\end{table}
}
In the table (in which for the moment we restrict ourselves to the
classical series) we use the notation of \cite{brade90,macka94}.
The Coxeter and dual Coxeter numbers of the algebra are $h$ and
$h^\vee$, whilst the difference between them and the $h',h'^\vee$
is well-known (see {\em e.g.\ }\cite{dorey93}) to be incapable of
absorption into a single overall convention; $t={h'\over h}{h^\vee
\over h'^\vee}$. A general form for $x$ and $q$ which gives the
correct soliton mass ratios is then
\be\label{xq}
x(\theta) = e^{h\lambda\theta}
\;,\hspace{0.3in}
q=e^{\omega i\pi}\;,
\ee
with
$$ \lambda = {4\pi\over\beta^2} - {h^\vee\over h}\;, \hspace{0.3in}
\omega={h\over h'} \left({4\pi\over \beta^2}-t\right)\;.
$$
The crossing parameter $x(i\pi)$ is expected to be $+q^{h'}$ for algebras
whose affine dual is untwisted, and $-q^{h'}$ for those whose affine
dual is twisted (see \cite{artz95,jimbo86}).
This works fine for all except the $a^{(2)}$ series, where we obtain
minus the expected value. This may be resolved by multiplying
$\la$ and $\om$ by 2, since neither crossing nor any other pole knows
about the scale of the exponent in $x$ and $q$ (this ambiguity is
mentioned in \cite{deliu95}).
However, such an {\em ad hoc} resolution is clearly unsatisfactory,
particularly since the form (\ref{xq}) of $x$ is by now well-known
\cite{babic94,brack93,efthi93,smirn91} and is due to the
transformation from the physical `spin' gradation to the homogeneous
gradation, in which all the $\theta$-dependence is transferred to
the step operator corresponding to $\al_0$, and in which the
$R$-matrices (\ref{specdecom}) are written.
This is the generalisation to the
nonsimply-laced case of the redefinition of in-/out-states (relative
to the quantum group representations) required in \cite{gande95}.
Specifically, there exist non-local charges which form
$U_q(\hat{g}^\vee)$ \cite{babic94,berna90,efthi93,feld91}
and which transform as
$$
L_\theta(E^\pm_i) = e^{\pm s_i\theta}E^\pm_i
\hspace{0.5in}{\rm where}\;\;\;
s_i = {8\pi\over \al_i^2\bt^2}-1
$$
under the action of a Lorentz boost $L_\theta$.
For any $y$ we now have
\be
y^{h_i} E^{\pm'}_j y^{-h_i}  =  y^{\pm\al_i.\al_j}x
 E^{\pm'}_j
\ee
where the $E^{\pm'}_i$ are step operators, and the $h_i$ Cartan
subalgebra generators, in any gradation.
If we do a similarity transformation on the
(homogeneous gradation) $R$-matrix
$$
R(x,q) \rightarrow \tau_{21} R(x,q) \tau_{12}^{-1} \;,
$$
where
$$
\tau_{12} = y^{-h_i a_i}(\theta_1) \otimes y^{-h_i a_i}(\theta_2)
$$
(for some unspecified $y(\theta)$), then we must solve
$$
y^{\al_i.\al_j\,a_j} x^{\delta_{i0}} = e^{s_i\theta}
$$
if we wish to obtain the homogeneous gradation.
In fact we can solve for $x$ without needing to know $y$ or $a_j$
by taking  the product of the $n_i$th powers of these equations
(where $n_i$ are the Kac marks of $\hat{g}^\vee$),
and then have
\beaa
x  & = &
\exp \sum_{i=0}^r n_i \left(
{8\pi\over\beta^2\alpha_i^2}-1\right)\theta
\\[0.1in]
& = & \exp h \lambda\theta \;,
\eeaa
where
$$
\lambda=\left({4\pi\over\beta^2} - {h^\vee\over h}\right)\;.
$$

It is now clear how flexible pole structure occurs: for self-dual
algebras, with $h=h^\vee$, the $\beta$-dependences of $x$ and $q$ are
proportional, and the `base' pole (corresponding to an unexcited soliton)
does not involve $\beta$: masses do not renormalise.
For the other algebras we have $x(\theta_r)=q^r$ at
$$
\theta_r = i\pi {r \over h'}{ 1- {\beta^2\over 4\pi}t \over
1-{\beta^2\over 4\pi}{h^\vee\over h}} \;.
$$
(We write out $\lambda$ and $\omega$ explicitly
each time for transparency.)
If $M_a \sim \sin\left({a\pi\over H}\right)$,  the
fusions $a,b\rightarrow a\pm b$ at $r=a\pm b$ (in the direct and
crossed channels respectively) require
$$
H= h' {1-{\beta^2\over 4\pi}{h^\vee\over h} \over
1- {\beta^2\over 4\pi}t} \;.
$$
This gives precisely the particle mass ratios: {\em i.e.\ }if
exact $S$-matrices exist in this form, the soliton mass ratios
for nonsimply-laced untwisted algebras must
be as in \cite{deliu94b} and the note to \cite{macka94}.

We can interpolate between affine dual algebras in two ways:
either by \cite{deliu92}
$$
H = h' + {h'\over h} (th-h^\vee) {{\beta^2\over 4\pi} \over
 1- {\beta^2\over 4\pi}t} \;,
$$
or by \cite{dorey93}
$$
{1\over H} = {1\over h'} + {1\over h'h} (h^\vee-th)
{{\beta^2\over 4\pi} \over 1-{\beta^2\over 4\pi}{h^\vee\over h}} \;.
$$
(Note that this does not work so neatly if we try to interpolate
$h$ and $h^\vee$ rather than $h'$ and $h'^\vee$.)
Notice how duality might work:
$\beta\mapsto 4\pi/\beta$ leads to
$$
H\mapsto h'^\vee
{1-{\beta^2\over 4\pi}{h\over h^\vee} \over
1- {\beta^2\over 4\pi}{1\over t}}\;,
$$
so that $g\mapsto g^\vee$ (with $(h,h^\vee)\mapsto(h^\vee,h)$
and $t\mapsto 1/t$). However, the exponents $\la$ and $\om$
receive an overall factor $-\bt^2/4\pi$, so that the
$S$-matrix is fundamentally different (with, for example,
no excited state poles); any duality must be much more subtle
than that for the particles.

Now consider poles for breathers and excited solitons.
The possibility of such objects arises because of
the ambiguity in the above pole:
$x(\theta^{(a+b)}_p)=q^{a+b}e^{-2pi\pi}$ gives
$$
\theta^{(a+b)}_p = {i\pi\over H}\left(a+b -{2p\over \omega}\right) \;,
$$
which gives, via the usual mass relation, physical poles for
$p=1,2,...\leq\frac12\omega(a+b)$, all transforming at the singular value
of the $R$-matrix $x=q^r$ and thus in the same particle multiplet.
Of course, the possibility of such poles does not necessarily lead
to their presence in the $S$-matrix, and, as we saw for $d_3^{(2)}$,
such poles do not all correspond to excited solitons.

For scalar breathers we need
$x(\theta^{(0)}_p)=q^{h'}e^{-2pi\pi}e^{-(h^\vee-th)i\pi}$.
This occurs at
$$
\theta_p^{(0)} = i\pi\left( 1 - {2p\over h \lambda} \right) \;,
$$
so that if a parent soliton of mass $M_a$ has a pole at $i\pi$
in its self-interaction then it can have a spectrum of breathers
with masses
\be\label{brmassform}
m_p^{(a)} = 2M_a\sin\left(
{p\pi\over\kappa h\lambda}\right) \hspace{0.2in}(\mbox{for}\hspace{0.2in}
p=1,2,...\leq \frac12 h\lambda)
\;.
\ee
Note that the running couplings $\lambda$ and $\omega$ become zero
at $\bt^2=4\pi {h \over h^\vee}$ and $4\pi/t$ respectively, and
the excited states disappear from the spectrum for large $\bt$.
In the simply-laced cases, and by analogy with the sine-Gordon theory,
these coincide to give a rational limit for the $R$-matrix and
an expected Yangian symmetry as $\bt^2\rightarrow 4\pi$.
In the non-simply-laced theories, we do not have $x,q\rightarrow 1$
simultaneously. The nature of these limits should be studied:
for example, at $\bt^2=4\pi/t$ we appear to have a trivial
$S$-matrix and thus a free theory. Recall also that at the point
at which the $n$th breather enters the spectrum the sine-Gordon
$S$-matrix becomes reflectionless and equal to the $d_n^{(1)}$
real ATFT $S$-matrix \cite{brade90}. It seems to us that all such
special values of $\la$ and $\om$ are potentially interesting and
deserving of study.

\newpage

\section{Crossing symmetry of $\cR_{1,1}$}\label{cross}

\vspace{.3cm}
In order to prove the crossing symmetry of the R-matrix we use
the Birman-Wenzl-Murakami algebra \cite{birman87}
in its diagrammatic notation (for
details see \cite{macka92b}):
\vspace{1cm} \\
Identity $\equiv$ \identity \hspace{1cm}
Monoid $\equiv$  \monoid \vspace{.5cm} \\
Braid $\equiv$ \braid \hspace{1.3cm} (Braid)$^{-1}$ $\equiv$
\antibraid .
\vspace{.5cm} \\
These symbols can be multiplied simply by concatenation and they
satisfy the following relations: \vspace{.3cm}
\be
\braid - \antibraid = m\left( \identity - \monoid \right) \label{bwmrel}
\ee
\vspace{0.3cm}
\be
\monoid\braid = l^{-1} \monoid, \hspace{1.5cm} \antibraid\monoid =
l\monoid
\ee
\vspace{.3cm}\\
in which $m = q-q^{-1}$ and $l = -q^{2n+1}$ in the case of
$c_n^{(1)}$.
Now we want to express the lowest R-matrix in terms of these symbols.
We know from (\ref{specdecom}) that:
\be
\cR^{(TPG)}_{1,1}(\th) = \cP_{2\la_1} + \lag 2\rag \cP_{\la_2} + \lag
2n+2\rag \cP_0 \;.
\ee
The projectors can be expressed in the following way \cite{macka92b}:
\bea
\nn \\
\cP_{2\la_1} &=& \frac1{1+q^2} \biggl[ \identity + q\braid +
\frac{q-q^{-1}}{q^{-1}+q^{2n+1}} \monoid \biggr] \nn \\ \nn \\
\cP_{\la_2} &=& \frac1{1+q^{-2}} \biggl[ \identity - q^{-1} \braid +
\frac{(1+q^{-2n-2})(q-q^{-1})}{(1-q^{-2n})(q^{-1}+q^{2n+1})} \monoid
\biggr] \nn \\ \nn \\
\cP_0 &=& -\frac{q-q^{-1}}{(1-q^{-2n})(q^{-1}+q^{2n+1})} \monoid \;.
\label{projectors}
\eea
\vspace{0.3cm} \\
Using the expressions (\ref{projectors}) in the expression of
$\cR_{1,1}^{(TPG)}$ we obtain after some algebra:
\be
\cR_{1,1}^{(TPG)}(\th) = \frac{x(1-q^2)}{x-q^2} \identity +
\frac{(x-1)q}{x-q^2} \braid +
\frac{x(x-1)(1-q^2)}{(x-q^2)(q^{2n+2}-x)} \monoid \;. \label{Rbwm}
\ee
We want to calculate the crossed version of this R-matrix:
\[\cR_{1,1}^{(TPG)cross}(i\pi-\th) =
[\si\cR_{1,1}^{(TPG)}(i\pi-\th)]^{t_2}\si\;. \]
Now the reason for the use of the BWM algebra becomes clear, since
under the change to the crossed version the generators of the BWM
algebra are transformed simply by turning them through $90$ degrees.
This exchanges the monoid and identity operators and the braid
operator changes to a (braid)$^{-1}$ operator.
Noting also that $x(i\pi-\th) = x^{-1}(\th)q^{2(n+1)}$ we obtain:
\bea
\cR_{1,1}^{(TPG)cross}(i\pi-\th) &=&
\frac{x^{-1}q^{2n+2}(1-q^2)}{x^{-1}q^{2n+2}-q^2} \monoid +
\frac{(x^{-1}q^{2n+2}-1)q}{x^{-1}q^{2n+2}-q^2} \antibraid \nn \\ \nn
\\ && + \frac{x^{-1}q^{2n+2}(x^{-1}q^{2n+2}-1)(1-q^2)}
{(x^{-1}q^{2n+2}-q^2)(q^{2n+2}-x^{-1}q^{2n+2})} \identity \;.
\label{Rcrbwm}
\eea
Using relation (\ref{bwmrel}) we can rewrite this in the following
form:
\bea
\cR_{1,1}^{(TPG)cross}(i\pi-\th) &=&
\frac{x(1-q^2)(x-q^{2n+2})}{(q^{2n+2}-xq^2)(1-x)}
\identity +
\frac{q^{2n+2}-x}{q^{2n+1}-xq} \braid \nn \\ \nn \\
&& + \frac{x(1-q^2)}{q^{2n+2}-xq^2} \monoid \;.
\eea
Comparing this with the expression (\ref{Rbwm}) we can see:
\be
\cR_{1,1}^{(TPG)cross}(i\pi-\th) =
\frac{(x-q^2)(x-q^{2n+2})}{(1-x)(q^{2n+2}-xq^2)}
\cR_{1,1}^{(TPG)}(\th) \;.
\ee
Writing $x$ and $q$ in terms $\mu$ and $\om$ as in (\ref{xqdef}) we can
write the scalar factor
\[
\frac{(x-q^2)(x-q^{2n+2})}{(1-x)(q^{2n+2}-xq^2)} =
\frac{\sin(\pi(\mu-\om)) \sin(\pi(\mu-(n+1)\om))}{\sin(\pi(-\mu+n\om))
\sin(\pi(-\mu)} \;,
\]
in which the denominator of the factor on the right hand side is equal
to its numerator
under the transformation $\th \to i\pi-\th$ and therefore we arrive at
the desired result:
\be\label{cr11}
c_{1,1}(i\pi-\th)\cR_{1,1}^{(TPG)cross}(i\pi-\th) =
c_{1,1}(\th)\cR_{1,1}^{(TPG)}(\th) \;,
\ee
in which
\be\label{c11}
c_{1,1}(\th) = \sin(\pi(\mu-\om))\sin(\pi(\mu-(n+1)\om)) \;.
\ee

\section{Some formulae concerning
$F_{a,b}(\mu)$}\label{scalar}

In this appendix we will derive some important formulae for the scalar
factor $F_{a,b}(\mu)$, introduced in section 3.2, which have been used
to calculate the S-matrix elements for the scattering of soliton bound
states in section 3.3.

The scalar factor $F_{1,1}(\mu)$ has been given explicitly in equation
(\ref{F11}) in terms of gamma functions.
In order to construct
breather S-matrices we need the following two identities:
\bea
F_{1,1}(\mu+\frac{n+1}2\om)F_{1,1}(\mu-\frac{n+1}2\om) =
\frac{\sin(\pi(\mu+\frac n2\om-\fo)) \sin(\pi(\mu-\frac
n2\om-\fo))}{\sin(\pi(\mu-\frac n2\om+\fo))
\sin(\pi(\mu+\frac n2\om+\fo))} \times \nn \\
\times \bl\frac n2\om+\fo \br \bl\frac n2\om+\fo+1 \br
\bl-\frac n2\om+\fo \br \bl\frac32n\om+\fo+1\br \nn \\
\nn \\
F_{1,1}(\mu+p) = (-1)^p \prod_{l=1}^p \frac {\lb-\om+l-1\rb
\lb\om+l\rb \lb-(n+1)\om+l-1\rb \lb(n+1)\om+l\rb}{\lb n\om+l\rb \lb
(n+2)\om+l+1 \rb \lb l\rb \lb l-1\rb} F_{1,1}(\mu) \nn \\
(\mbox{for any integer }p).
\eea
In the last expression we have used the additional notation $\lb y\rb
\equiv \sin(\frac{\pi}{(n+1)\la}(\mu+y))$.
These formulas can be obtained from (\ref{F11}) by direct
calculation\footnote{In all these calculations one has to assume that
$\om$ is not an integer. This case would correspond to $q =$ root of
unity which should be examined separately}
using the fundamental
property of the gamma function $\Ga(z+1) = z\Ga(z)$ (for $z \neq
0,-1,-2,...$) and the expansion of sine into an infinite product.
(Similar calculations have also been performed in
\cite{gande95}.) Together with formula (\ref{Fab}) for the general
scalar factor $F_{a,b}$ we can derive the following
\bea
&&F_{a,b}(\mu+\frac{n+1}2\om)F_{a,b}(\mu-\frac{n+1}2\om) = \nn \\
&&= \prod_{j=1}^a \prod_{k=1}^b \frac{\sin(\pi(\mu+\fo(2j+2k-a-b-3+n)))
\sin(\pi(\mu+\fo(2j+2k-a-b-3-n)))} {\sin(\pi(\mu+\fo(2j+2k-a-b-1-n)))
\sin(\pi(\mu+\fo(2j+2k-a-b-1+n)))} \nn \\
& & \times \bl\fo(2j+2k-a-b-1+n)\br \bl\fo(2j+2k-a-b-1+n)+1\br \nn \\
& & \times \bl\fo(2j+2k-a-b-1-n)\br \bl\fo(2j+2k-a-b-1+3n)+1\br \;.
\eea
This last formula reduces to formula (\ref{fabfab}) in section 3.3 and
we see that the
infinite product of gamma functions has been reduced to just a finite
product of sine functions. From this formula it is relatively
straightforward to calculate all S-matrix elements for the scattering of
breathers with themselves and of solitons and excited solitons with
breathers. The results of these calculations were listed at the end of
section 3.3.

\vfill
\newpage

\section{Pole structure of rational $c_n$-invariant
$S$-matrices}\label{poles}

The rational $c_n$-invariant $S$-matrices are given in \cite{macka92},
their $R$-matrix structure being given by the TPG as in section 2,
but with $\lag r \rag$ replaced by $-[r]$ where
$$
[r] \equiv {\th+ {i\pi r\over h'} \over \th- {i\pi r\over h'}} \;,
$$
and $U_q(g)$-modules replaced by $g$-modules; and with the
`minimal' $S$-matrices ($R$-matrices made crossing symmetric and
unitary and without any physical poles) being multiplied by CDD
factors equal to the
na\"{\i}ve $d_{n+1}^{(2)}$ real ATFT $S$-matrix numerators of
\cite{brade90}. (These would describe
Gross-Neveu models. For principal chiral models we would take
two copies of the minimal $S$-matrix, acting on the left- and
right-global $g$-symmetry modules, and then multiply by a single CDD
factor. Thus the pole structure is different \cite{hollo94}, although
still explained by an analysis similar to that below.)
First we should point out that \cite{macka92} contains a mistake:
in the light of (\ref{fusion}) we can see that the right-hand side
of equation (12) of that paper, describing $S_{a,b}$ ($a\geq b$),
should be multiplied by
$$
\prod_{i=1}^b \prod_{j=1}^{a-1} [a-b+2i-2j] \;.
$$
For $a+b<n+1$ the rational $S_{a,b}$ has simple poles at
$(a+b){i\pi\over h'}$, corresponding to the direct-channel
fusing $a,b\rightarrow a+b$, and at $(h'-a+b){i\pi\over h'}$,
corresponding to the fusing $a,b\rightarrow a-b$, and their
crossed-channel partners.

There are also double poles at $a-b+2i\;$ ($i=1,..,b-1$)
and $h'+2j-a-b\;$ ($j=1,..,b-1$). These correspond
to diagrams of the form of figure 1,

\begin{center}
\begin{picture}(340,160)(-20,-30)
\put(0,0){\line(1,2){20}}
\put(120,0){\line(-1,2){20}}
\put(20,40){\line(0,1){60}}
\put(20,40){\line(4,3){80}}
\put(20,100){\line(4,-3){80}}
\put(100,40){\line(0,1){60}}
\put(20,100){\line(-1,2){20}}
\put(100,100){\line(1,2){20}}
\put(60,70){\circle*{8}}
\put(-5,-10){\shs{\fns{$b$}}}
\put(117,-10){\shs{\fns{$a$}}}
\put(117,142){\shs{\fns{$b$}}}
\put(-2,142){\shs{\fns{$a$}}}
\put(12,65){\shs{\scs{$i$}}}
\put(106,65){\shs{\scs{$i$}}}
\put(35,45){\shs{\scs{$b-i$}}}
\put(67,45){\shs{\scs{$a+i$}}}
\put(36,89){\shs{\scs{$a+i$}}}
\put(67,89){\shs{\scs{$b-i$}}}
\put(30,-30){\shs{\em Figure 6a}}

\put(200,0){\line(1,2){20}}
\put(320,0){\line(-1,2){20}}
\put(220,40){\line(0,1){60}}
\put(220,40){\line(4,3){80}}
\put(220,100){\line(4,-3){80}}
\put(300,40){\line(0,1){60}}
\put(220,100){\line(-1,2){20}}
\put(300,100){\line(1,2){20}}
\put(260,70){\circle*{8}}
\put(194,-10){\shs{\fns{$b$}}}
\put(318,-10){\shs{\fns{$a$}}}
\put(318,142){\shs{\fns{$b$}}}
\put(197,142){\shs{\fns{$a$}}}
\put(212,65){\shs{\scs{$j$}}}
\put(306,65){\shs{\scs{$j$}}}
\put(235,45){\shs{\scs{$b-j$}}}
\put(267,45){\shs{\scs{$a-j$}}}
\put(236,89){\shs{\scs{$a-j$}}}
\put(267,89){\shs{\scs{$b-j$}}}
\put(235,-30){\shs{\em Figure 6b}}

\end{picture}
\end{center}

\noindent In figure 6a the direct channel (left-to-right)
angle at the internal $S$-matrix is $(a-b){i\pi\over h'}$,
at which it has neither pole nor zero,
while in figure 6b the angle is $(h'-a-b){i\pi\over h'}$, at which
$S_{a-j,b-j}$, too, is finite; thus both diagrams give double poles.

For $a+b\geq n+1$ the situation is different. For
$i=1,..,n-a$ figure 6a still describes a double pole in the same way.
In figure 6b, we still have a double pole for $j=1,..,a+b-n-2$, but
for higher $j$ the internal $S_{a-j,b-j}$ has poles. The correct
orders of the poles in $S_{a,b}$, when the zeros arising from
the truncation of the TPG have been taken into account (as noted
in \cite{hollo94}), are two for $j=a+b-n-1$, and three for
$j=a+b-n,..,b-1$, and it is these that we must explain.
Now the internal $S_{a-j,b-j}$ has poles of order one and two
respectively, leading to na\"{\i}ve orders for these diagrams of three
and four. However, at these internal angles $S_{a-j,b-j}$
projects onto precisely that part of the $(a-j,b-j)$ TPG which is
lost in truncation of the $(a,b)$ TPG, giving us an extra
zero and the correct pole orders.
The only pole we are unable to explain satisfactorily (again
as pointed out in \cite{hollo94}) is that at ${i\pi\over 2}$
in $S_{a,n+1-a}$, a problem which does not arise either for the
principal chiral model $S$-matrix or in the
trigonometric case. This mechanism applied to the trigonometric case
also gives the correct pole orders for the $p=0$ poles but
fails to explain the higher ones, which we expect to require
diagrams of the form of figure 3 or of even higher order.

It is interesting to note that these diagrams plus the tree-level
diagrams now explain {\em all} the $R$-matrix singularities,
at which all the edges directed into a subgraph of the TPG
have the same value. Further, we can nest a series of diagrams
of the type of figures 6a and 6b, with, say, $i=j=1$, to give
all the singularities of the $R$-matrix in two diagrams,
corresponding to the horizontal and vertical links respectively
in the TPG.
It therefore becomes interesting to consider whether some method
along these lines might help us to understand the singularities of
other ({\em e.g.\ }$b_n-$ or $d_n-$invariant) $R$-matrices whose
decomposition is not known. There are intriguing similarities
with the singularity theorem
of Chari and Pressley \cite{chari91}, who pointed out that the
singularity at the crossing point of the $R$-matrix associated
with a Lie algebra $\tilde{g}$ must also be a singularity of
the $R$-matrix associated with $g$, where the Dynkin
diagram of $\tilde{g}$ is a sub-diagram of that of $g$. (Presumably this
result also extends to singularities other than that at the crossing
point.) This theorem has recently been used powerfully to
help understand the fusion properties of $U_q(\hat{g})$- and
Yangian representations \cite{chari95}.

\vspace{2cm}

\end{document}